\documentclass[fleqn,usenatbib,useAMS]{mnras}

\usepackage[T1]{fontenc}
\usepackage{ae,aecompl}

\usepackage{graphicx}	% Including figure files
\usepackage{amsmath}	% Advanced maths commands
\usepackage{amssymb}	% Extra maths symbols

\title[The magnetic field of HD\,54879]{
A short and sudden increase of the magnetic field strength
and the accompanying spectral variability in the O9.7\,V star HD\,54879
}

\author[Hubrig et al.\ 2018]{
S.~Hubrig$^{1}$\thanks{E-mail: shubrig@aip.de},
M.~K\"uker$^{1}$,
S.~P.~J\"arvinen$^{1}$,
A.~F.~Kholtygin$^{2,3}$,
M.~Sch\"oller$^{4}$, 
\and
E.~B.~Ryspaeva$^{2,5}$,
D.~D.~Sokoloff$^{6,7}$
\\
$^{1}$Leibniz-Institut f\"ur Astrophysik Potsdam (AIP), An der Sternwarte~16, 14482~Potsdam, Germany\\
$^{2}$ Saint-Petersburg State University, Universitetskij pr.~28, 198504 Saint-Petersburg, Russia\\
$^{3}$ Institute of Astronomy, Russian Academy of Sciences, ul. Pyatniskaya 48, Moscow 119017, Russia\\
$^{4}$European Southern Observatory, Karl-Schwarzschild-Str.~2, 85748 Garching, Germany\\
$^{5}$Main (Pulkovo) Astronomical Observatory, 196140 Saint-Petersburg, Russia\\
$^{6}$Department of Physics, Moscow State University, 119991 Moscow, Russia\\
$^{7}$IZMIRAN, Kaluzhskoe shosse, Troitsk, 108840 Moscow, Russia
}
\date{Accepted XXX. Received YYY; in original form ZZZ}

\pubyear{2018}

\begin{document}
\label{firstpage}
\pagerange{\pageref{firstpage}--\pageref{lastpage}}
\maketitle

% Abstract of the paper
\begin{abstract}
Only eleven O-type stars have been confirmed to possess large-scale organized
  magnetic fields. 
  The presence of a $-600$\,G longitudinal magnetic field in the O9.7\,V star
  HD\,54879 with a lower limit of the dipole
  strength of $\sim$2\,kG,  was discovered a few years ago in the framework of 
the ESO large program ``B-fields in OB stars''.
 Our FORS\,2 spectropolarimetric observations from
  2017 October~4 to 2018 February~21 reveal the presence  of short and long-term
  spectral variability and a gradual magnetic field decrease from about $-300$\,G down to about $-90$\,G.
  Surprisingly, we discover on the night of 2018 February~17 a sudden, short-term increase of the magnetic field strength
  and measure a longitudinal magnetic field of $-833$\,G. The inspection of the FORS\,2 spectrum acquired during 
the observed magnetic field increase 
indicates a very strong change in spectral appearance with a significantly
lower photospheric temperature and a decrease of the radial velocity by several 10\,km\,s$^{-1}$. 
Different scenarios are discussed in an attempt to interpret our observations.
The FORS\,2 radial velocity measurements indicate that HD\,54879 is a member of a 
long-period binary.
\end{abstract}

% Select between one and six entries from the list of approved keywords.
% Don't make up new ones.

%%%%%%%%%%%%%%%%%%%%%%%%%%%%%%%%%%%%%%%%%%%%%%%%%%

%%%%%%%%%%%%%%%%% BODY OF PAPER %%%%%%%%%%%%%%%%%%

\section{Introduction}
\label{sec:intro}

The origin of magnetic fields in massive stars is still under debate. It has been
argued that magnetic fields could be fossil relics of the fields that
were present in the interstellar medium from which the stars have formed
(e.g.\ \citealt{Moss2003}). This, however, gives no explanation for
the low fraction of stars that are found to be magnetic. Different scenarios to explain 
why only a subclass of early-type stars are observably magnetic in the framework
of the fossil field theory were discussed by \citet{Borra1982},
who also indicated the need for future hydromagnetic studies, such as dynamo theory and the 
related problems of magnetic stability.
 In contrast, the magnetic field in the Sun and stars on the lower main sequence that have outer 
convection zones or are fully convective is believed to be 
the result of a dynamo process (e.g.\ \citealt{Charbonneau2014}).
Though there is no consensus on the exact mechanism, the global rotation 
and its impact on the convective gas motions are vital for the generation of a large-scale, oscillating 
magnetic field.  The observed rotation-activity relation and the existence
of activity cycles support the idea that the magnetic fields of stars on the lower main sequence are 
dynamo-generated. As the observed field geometries among such stars strongly vary, it is possible that 
the type of dynamo mechanism is not the same for all stars. Notably, all known mechanisms involve rotating convection.

\citet{Cantiello2011a} suggested that a subsurface convection zone due to 
partial ionization in massive stars may be the 
source of a global magnetic field, winding up toroidally with stochastic buoyancy breakouts at the surface, 
causing corotating magnetic bright spots at the surface of the star.
Dynamos operating in geometrically thin layers  have been studied in the context of the solar dynamo, 
where the overshoot layer below the convection zone has been assumed to be the location of the dynamo.  
\citet{Moss1990} reported that the magnetic field geometry produced by such dynamos is generally more complicated 
than a dipole. The width of the toroidal field belts was found to be about the same as the depth of 
the overshoot layer and the observed solar butterfly diagram is therefore not correctly reproduced. 
This finding was confirmed by \citet{Ruediger1995}, who also pointed out that this type of dynamo 
produces   cycle periods that are substantially shorter than the solar activity cycle period of 22\,yr.

Alternatively, magnetic
fields in massive stars may be generated by strong binary interaction, i.e.\ in stellar
mergers, or during a mass transfer or common envelope evolution \citep{Tout2008}.
In these events, angular momentum surplus
creates strong differential rotation in one of the stars \citep{Petrovic2005}
or in the merger product --- a key ingredient for
the generation of a magnetic field. The binary interaction also leads to enhanced
abundances of hydrogen burning products, e.g.\ nitrogen, at the stellar
surface. The number of O-type stars with confirmed magnetic fields detected at a
3$\sigma$-level is currently only eleven (e.g.\ \citealt{Grunhut2017,Schoeller2017}), 
and approximately half of them appear to be nitrogen-rich (e.g.\ \citealt{Martins2012}).

The most straightforward explanation for a large scale magnetic field in a massive star is a fossil field, 
which however would have to be stationary except for a slow decay and have a rather simple geometry. 
Cyclic activity, on the other hand, requires a large-scale dynamo mechanism. A dynamo
 could exist in the core, but it is unclear how much the magnetic field generated there could 
affect the stellar surface and surroundings. A nonaxisymmetric steady field produced by a dynamo 
in the core could appear to vary with time at the surface if the envelope undergoes cyclic changes. 
Torsional oscillations could cause a variation of the surface rotation and thus the observed cycle 
period.
However, the torsional oscillations would have to be driven by a force other than the Lorentz force and they would not affect the field strength.
MHD simulations of magnetic field generation in the core of A-type 
stars found rather complex and varying field geometries \citep{Brun2005}.  Because of 
the comparatively  high electric conductivity of the radiative envelope, a magnetic field generated 
in the core can not reach the surface through diffusion. However, instabilities involving buoyancy 
or shear could cause the rise of magnetic flux and therefore a variability of the surface 
magnetic field on a time scale comparable to that of the dynamo. However, the geometry of the surface 
field would then be much more complicated than a simple dipole. 
On the other hand, turbulent diffusion will destroy a possible fossil field, which means that a fossil 
field will be expelled from the core by convection. \citet{Featherstone2009} have studied the 
interaction between a fossil field and a core dynamo in A-type stars. They found that in the 
presence of a fossil field the core dynamo can generate stronger fields, thus raising the chance that 
buoyancy could transport some of the generated magnetic flux to the surface.

Apart from the O9.7\,V star HD\,54879, all previously studied magnetic O-type 
stars showed  spectral line variability with a period identified as the rotation
period. Of the known confirmed eleven magnetic O-type stars, five are associated
with the peculiar spectral classification Of?p. This classification was
introduced by \citet{Walborn1972} according to the presence of
C\,{\sc iii} 4650\,\AA{} emission with a strength comparable to the
neighbouring N\,{\sc iii} lines. Such stars are known to exhibit recurrent,
and apparently periodic, spectral variations in Balmer, He\,{\sc i},
C\,{\sc iii}, and Si\,{\sc iii} lines, sharp emission or P\,Cygni profiles in
He\,{\sc i} and the Balmer lines, and strong C\,{\sc iii} emission around
4650\,\AA{}. 
The detection of a magnetic field in HD\,54879 was achieved by \citet{Castro2015}
using the FOcal Reducer low dispersion Spectrograph (FORS\,2; \citealt{Appenzeller1998})
and the High Accuracy Radial 
velocity Planet Searcher in polarimetric mode (HARPS\-pol) in the framework of the ESO Large Prg.~191.D-0255.
The authors reported the presence of a $-600$\,G longitudinal magnetic
field with a lower limit of the dipole strength of $\sim2$\,kG. Magnetospheric parameters 
of HD\,54879 were recently presented by \citet{Shenar2017},
who also suggested a rotation period of $\sim5$\,yr.
Noteworthy, among the single magnetic O-type stars, HD\,54879 has the second
strongest magnetic field after the Of?p star NGC\,1624-2, for which a dipole
strength of $\sim$20\,kG was estimated by \citet{Wade2012}. The stars HD\,54879 and NGC\,1624-2
also show the lowest $v \sin i$ values, lower than 8\,km\,s$^{-1}$. Current
studies of O-type stars indicate that their magnetic fields are dominated by
dipolar fields tilted with respect to the rotation axis (the so-called
oblique dipole rotator model). Thus, the rotation period is frequently
determined from studying the periodicity in the available magnetic data. As
of today, an estimation of the rotational periodicity using magnetic field
measurements and spectral variability was performed for ten magnetic O-type
stars, indicating rotation periods between 7\,d for HD\,148937 \citep{Naze2008} and 
several decades for HD\,108 \citep{Naze2001}, while most normal O-type stars have typical rotation
periods in the range from 2 to 4\,d. In contrast, for HD\,54879 only very few magnetic field measurements,
three measurements using FORS\,2 and three measurements using
HARPS\-pol were available before we started the magnetic field monitoring using FORS\,2 observations
within the framework of the programme 100.D-0110(A).
In the following, we report on our results obtained from the polarimetric observations 
of HD\,54879 carried out using FORS\,2 in spectropolarimetric mode.

%----------------------------------------------------------------------------

\section{Observations and magnetic field analysis}
\label{sec:obs}

\begin{table}
\centering
\caption{Longitudinal magnetic field values of HD\,54879 from
 FORS\,2 data. The S/N is measured at 4800\,\AA{}.
}
\label{tab:FORS}
\begin{tabular}{ccr@{$\pm$}lr@{$\pm$}lr@{$\pm$}lc}
\hline
\multicolumn{1}{c}{MJD} &
\multicolumn{1}{c}{S/N} &
\multicolumn{2}{c}{$\left<B_{\rm z}\right>_{\rm all}$} &
\multicolumn{2}{c}{$\left<B_{\rm z}\right>_{\rm hyd}$} &
\multicolumn{2}{c}{$\left<B_{\rm z}\right>_{\rm N}$} \\
&
&
\multicolumn{2}{c}{(G)} &
\multicolumn{2}{c}{(G)} &
\multicolumn{2}{c}{(G)}
\\
\hline
56696.2341$^1$& 2360 & $-$460 & 65 & $-$639 & 121 & 76 & 66   \\
56697.2162$^1$& 2400 & $-$521 & 62 & $-$877 &  91 & 23 & 63  \\
57099.0150$^1$& 3060 & $-$527 & 45 & $-$633 &  65 & 52 & 45  \\
58030.2744    & 1700 & $-$242 &120 & $-$313 & 158 &$-$33 & 107 \\
58031.3448    & 1870 & $-$51  &86  & $-$133 & 117 &$-$22 & 70 \\
58033.3604    & 2160 & $-$154 & 74 & $-$138 & 118 & 45& 68 \\
58035.2867    & 1710 & $-$261 & 83 & $-$522 & 139 & 18& 77 \\
58036.3591    & 1780 & $-$245 & 65 & $-$434 & 129 &$-$54& 73  \\
58041.3450    & 3130 & $-$187 & 57 & $-$126 &  91 &$-$34& 55  \\
58043.3604    & 1660 & $-$125 & 55 &     33 &  94 & $-$35&55 \\
58046.3496    & 2080 & $-$248 & 86 & $-$141 & 152 & $-$80& 76  \\
58049.2865    & 2290 & $-$136 & 62 & $-$201 & 117 & 11& 59 \\
58051.3598    & 3970 &  $-$16 & 82 & $-$223 & 104 & $-$62& 90 \\
58062.2035    & 2240 & $-$186 & 55 & $-$286 &  98 & 66& 58 \\
58073.3423    & 2410 & $-$202 & 52 & $-$260 &  81 & 13& 56 \\
58098.3371    & 2090 &  $-$58 & 53 & $-$163 & 113 & $-$12& 58 \\
58099.3109    & 2300 & $-$153 & 66 & $-$196 & 104 & $-$21& 62 \\
58100.3289    & 2010 &  $-$75 & 78 & $-$138 & 124 & $-$22& 72 \\
58103.3134    & 2640 & $-$218 & 52 & $-$242 & 105 & $-$17& 54 \\
58149.2310    & 2240 & $-$129 & 54 & $-$160 &  98 & 57& 53  \\
58150.0603    & 1740 & $-$101 & 62 & $-$137 & 122 & 60& 63 \\
58151.1108    & 2380 & $-$113 & 60 &  $-$75 & 101 & 38& 56 \\
58152.1369    & 2950 & $-$195 & 57 & $-$128 & 102 & 86& 69 \\
58154.2698    & 1370 & $-$123 & 99 &  $-$89 & 128 & 62& 96 \\
58161.1319    & 2080 &  $-$36 & 56 &  $-$19 & 116 & 62& 75\\
58162.0188    & 2200 &  $-$85 & 59 & $-$245 & 112 &$-$79& 63 \\
58164.2586    & 1500 & $-$202 &132 & $-$114 & 166 &   85  & 158 \\
58166.0420    & 1130 & $-$833 &118 & $-$623 & 155 &$-$31 & 166 \\
58170.0332    & 2390 &  $-$91 & 49 &  $-$13 &  94 &$-$68 & 53 \\
\hline
\end{tabular}
\flushleft
{\bf Note:}\\
$^1${The measurements at MJD\,56696.2345 to MJD\,57099.0150
were previously reported by \citet{Castro2015}
and \citet{Schoeller2017}}.
\end{table}

The 26 new spectropolarimetric observations of HD\,54879  with FORS\,2
summarized in Table~\ref{tab:FORS} were obtained  between 2017
October 4 and 2018 February 21. The first two Columns list the modified Julian dates
(MJD) for the middle of the exposure and the signal-to-noise ratio (S/N) of the spectra. We used 
the GRISM 600B and the narrowest available slit width of 0$\farcs$4 to obtain
a spectral resolving power of $R\approx2000$. The observed spectral range from
3250 to 6215\,\AA{} includes all Balmer lines, apart from H$\alpha$, and
numerous helium lines. Further, in our observations we used a non-standard
readout mode with low gain (200kHz,1$\times$1,low), which provides a broader
dynamic range, hence allowing us to reach a higher S/N
in the individual spectra. The position angle of the retarder waveplate was
changed from $+45^{\circ}$ to $-45^{\circ}$ and vice versa every second exposure,
i.e.\ we have executed the sequence
$+45^{\circ}$$-45^{\circ}$$-45^{\circ}$$+45^{\circ}$
four times.
Using this sequence of the
retarder waveplate ensures an optimum removal of instrumental polarization.
The exposure time for the observations of the subexposures
$+45^{\circ}$$-45^{\circ}$$-45^{\circ}$$+45^{\circ}$
including overheads, accounted for about 6\,min.

A description of the assessment of the presence of a longitudinal magnetic
field using FORS\,1/2 spectropolarimetric observations was presented in our
previous work \citep[e.g.][and references therein]{Hubrig2004a, Hubrig2004b}.
Improvements to the methods used, including $V/I$ spectral rectification
and clipping, were detailed by \citet{Hubrig2014}.

\begin{figure}
 \centering 
        \includegraphics[width=0.95\columnwidth]{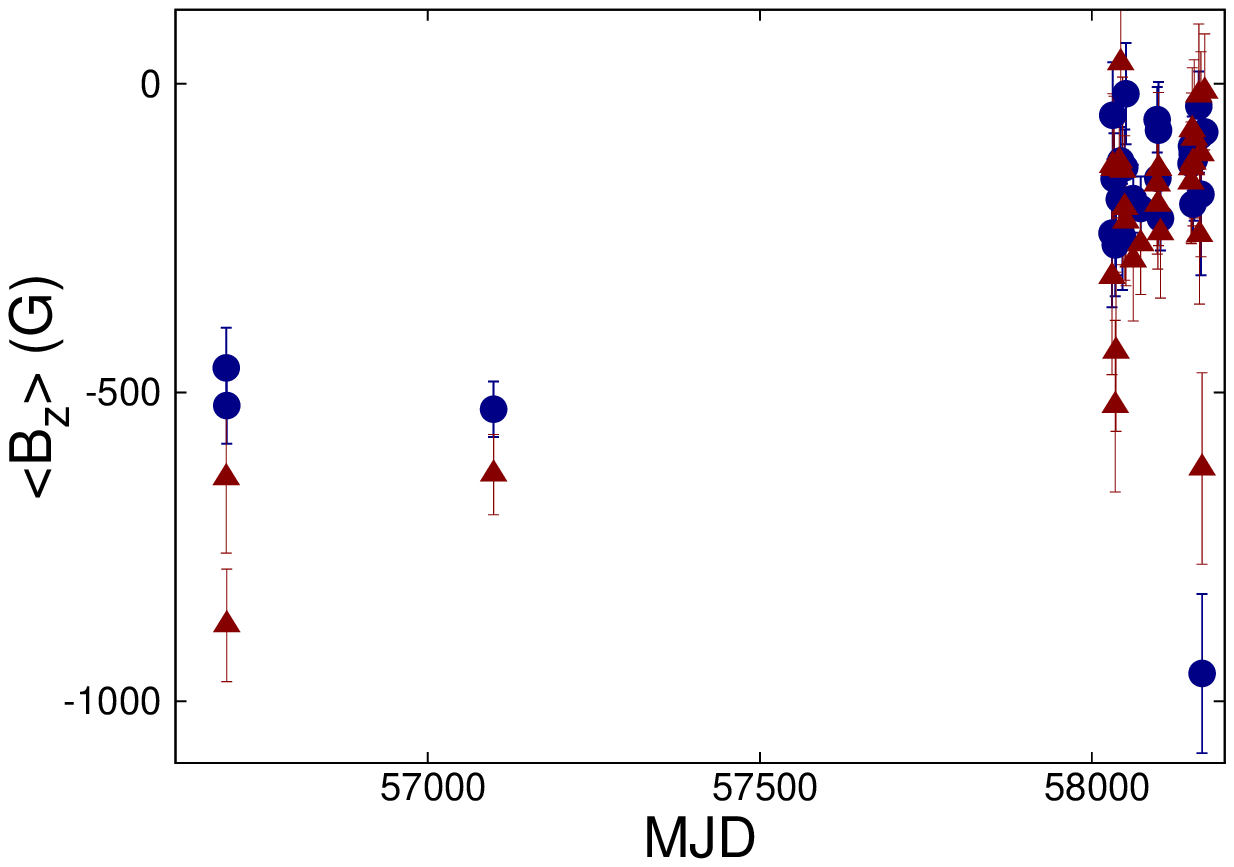}
        \includegraphics[width=0.95\columnwidth]{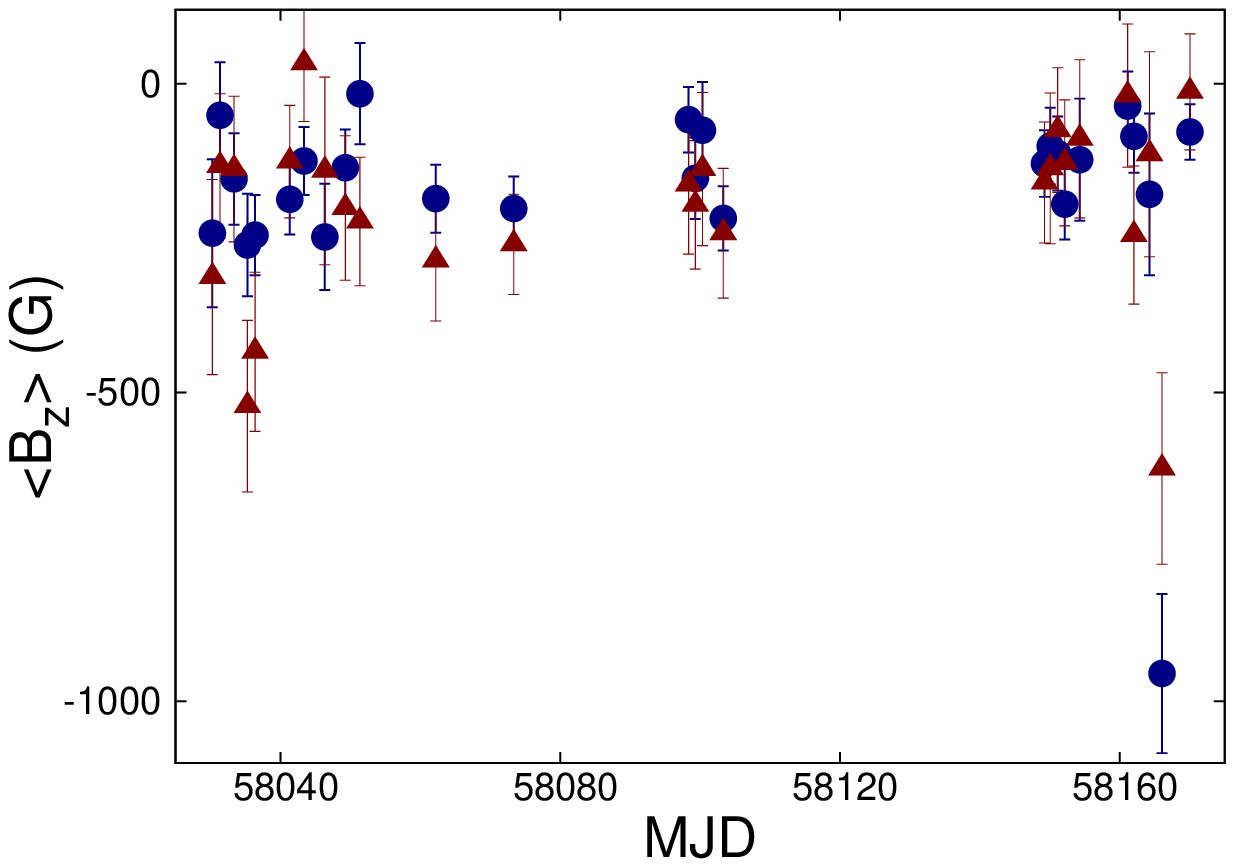}
                \caption{
{\it Upper panel:} Distribution of the mean longitudinal magnetic field values of
   HD\,54879 using the entire spectrum (blue circles) and those using only the hydrogen lines (red triangles) 
as a function of MJD between 2014 and 2018.
{\it Lower panel:}  Distribution of the mean longitudinal 
magnetic field values of HD\,54879 as a function of MJD between 2017 and 2018.
         }
   \label{fig:Bevol}
\end{figure}

The longitudinal magnetic field was measured in two ways: using the entire
spectrum including all available lines, or using exclusively hydrogen lines.
Furthermore, we have carried out Monte Carlo bootstrapping tests.
These are most often applied with the purpose of deriving robust estimates of standard
errors (e.g.\ \citealt{Steffen2014}). 
The measurement uncertainties obtained before and after the Monte Carlo
bootstrapping tests were found to be in close agreement, indicating the
absence of reduction flaws. The results of our magnetic field measurements,
those for the entire spectrum or only for the hydrogen lines are presented 
in Table~\ref{tab:FORS} in Columns~3 and 4, followed by the measurements using all lines
in the null spectra. The distribution of the mean longitudinal
magnetic field values as a function of MJD is
presented in Fig.~\ref{fig:Bevol}.

\begin{figure*}
 \centering 
        \includegraphics[width=0.33\textwidth]{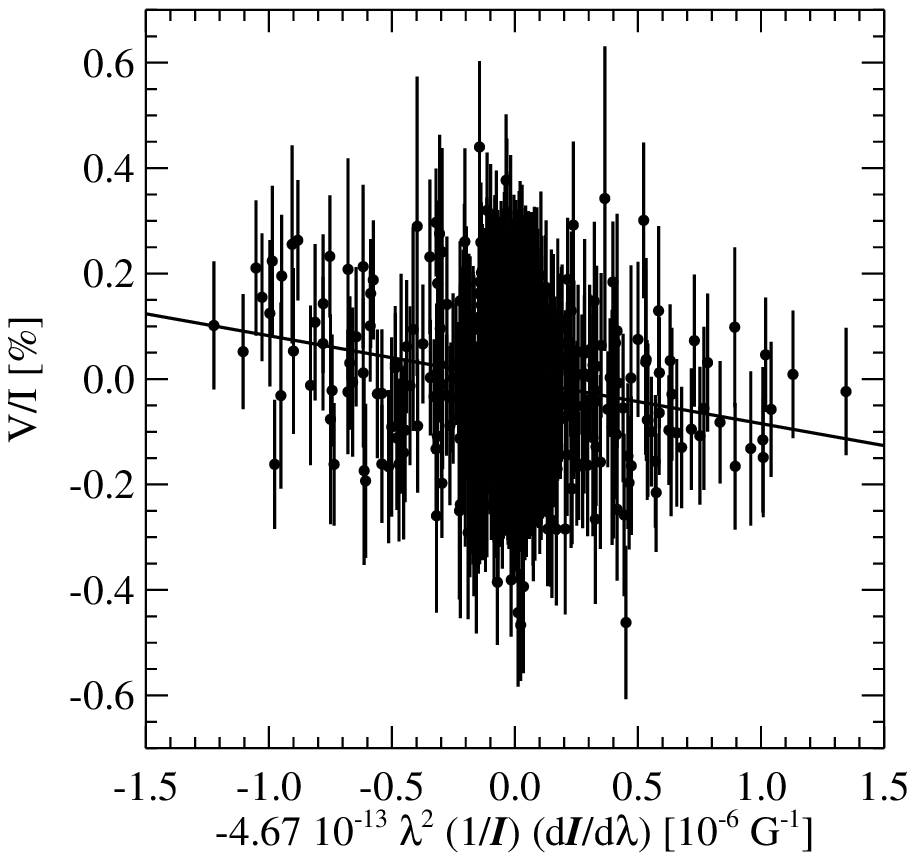}
  \includegraphics[width=0.33\textwidth]{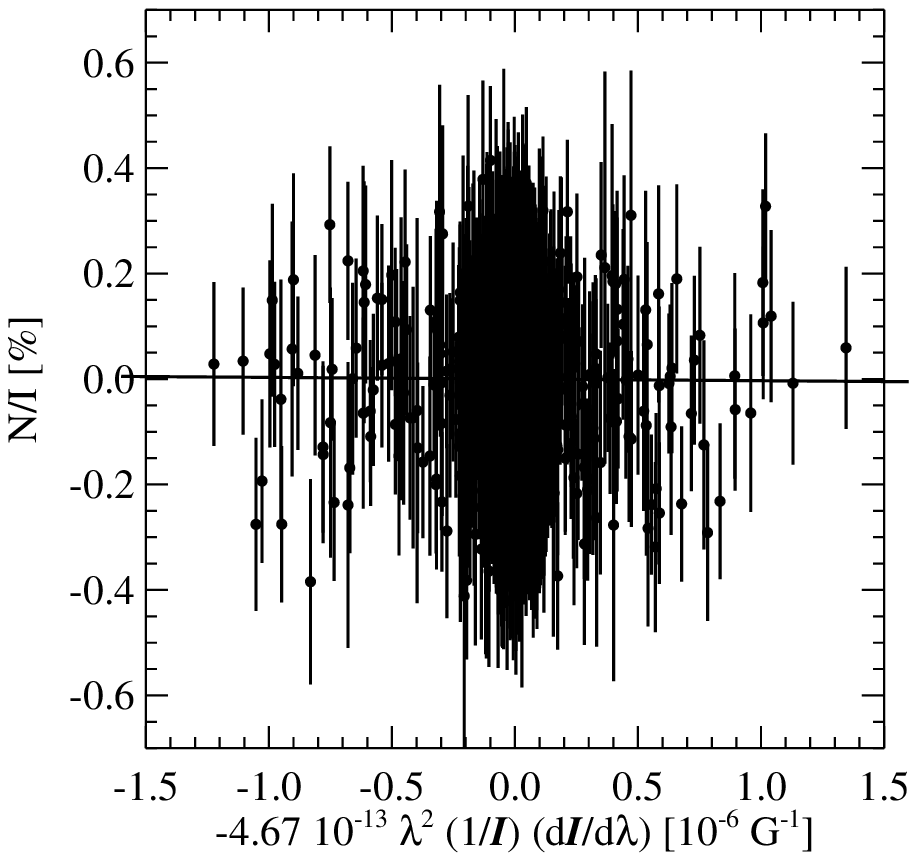}
  \includegraphics[width=0.33\textwidth]{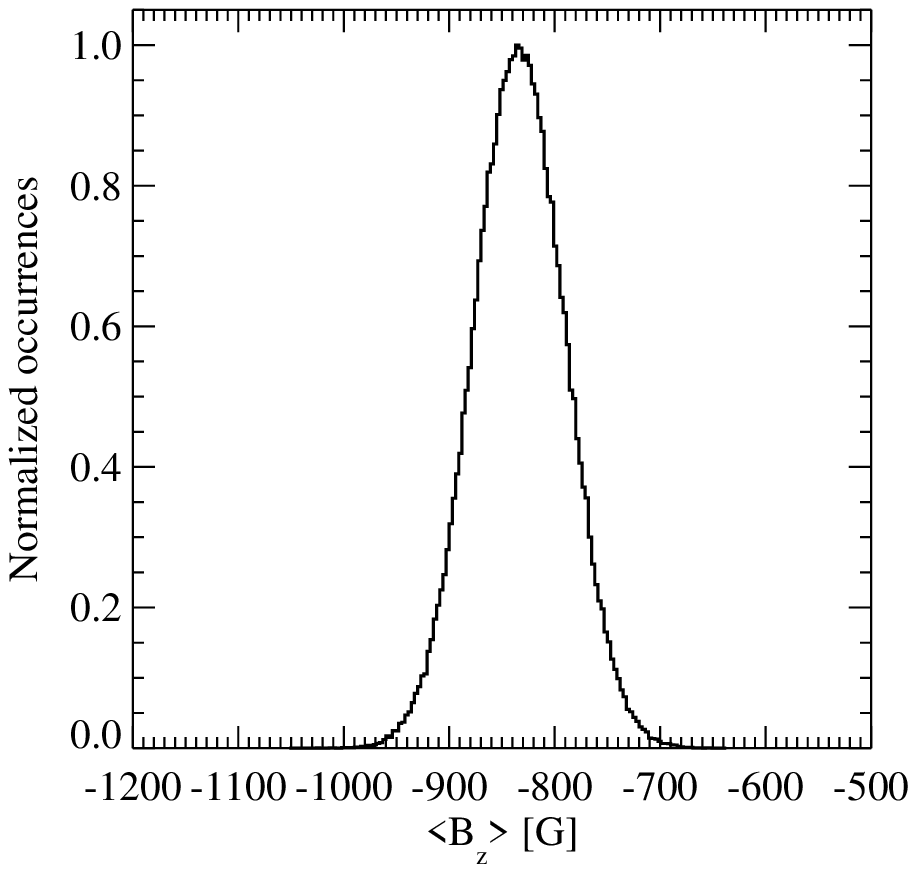}
\caption{
Regression analysis  of the FORS\,2 data of HD\,54879 obtained during the night of 2018 February 17 
considering the entire spectrum. {\it Left panel:} linear fit to Stokes~$V$. {\it Middle panel:} linear fit to the
$N$ spectrum. {\it Right panel:} distribution of the longitudinal magnetic field values $P(\left<B_{\rm z}\right>)$,
which were obtained via bootstrapping.
From the distribution $P(\left<B_{\rm z}\right>)$, we obtain the most likely value for the 
longitudinal magnetic field $\left< B_{\rm z} \right>=-833\pm118$\,G.
         }
   \label{fig:regr}
\end{figure*}

\begin{figure}
 \centering 
        \includegraphics[width=0.95\columnwidth]{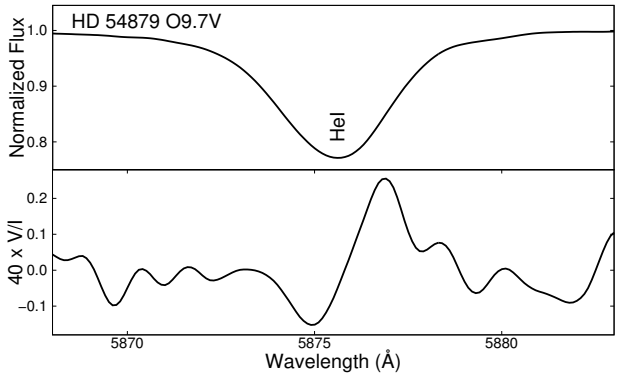}
        \caption{
{\it Top:}
Stokes~$I$ profile of \ion{He}{i}~5876.
{\it Bottom:}
A typical Zeeman feature in the Stokes~$V$ profile of \ion{He}{i}~5876 recorded on
2018 February 17.
         }
   \label{fig:zeeman}
\end{figure}

While the few previous observations in 2014/2015 indicated a
longitudinal magnetic field strength of the order of $-$500\,G to $-$600\,G,
our measurements using the observations of HD\,54879 carried out from 2017
October 4 to 2018 February 21 show that we approach the rotational phase with the best visibility of the 
magnetic equator. Over the four and a half months, the magnetic field was gradually decreasing, from
$\left<B_{\rm z}\right>\approx-300$\,G measured on 2017 October 4 down to
$\left<B_{\rm z}\right>\approx-90$\,G measured on 2018 February 21.
Surprisingly, the observations obtained on 2018 February 17 showed a
significant increase of the magnetic field strength, reaching
$\left< B_{\rm z} \right>=-833\pm118$\,G in the measurements using the entire spectrum and 
$\left< B_{\rm z} \right>=-623\pm155$\,G using exclusively the hydrogen lines. 
In Fig.~\ref{fig:regr} we present the results of our analysis 
of the FORS\,2 data of HD\,54879 obtained during the night of 2018 February 17 
considering the entire spectrum.
Further, as an example, we present in Fig.~\ref{fig:zeeman} 
a typical Zeeman feature in the Stokes~$V$ profile of \ion{He}{i}~5876 recorded on
2018 February 17.

Four nights later, on 2018 February 21, the magnetic field strength became again very weak, of the order of $-$90\,G. The
field decrease observed by us over about 4.5 months suggests that the
magnetic/rotation period is rather long, amounting to at least several
months, or even years. However, the sudden short-term field strength increase
observed on the night of 2018 February 17 (MJD\,58166.04) needs a
plausible explanation.

\section{Spectral changes accompanying the sudden increase of the magnetic field}

\citet{Jarvinen2017} studied the spectral variability of HD\,54879 on different timescales using all 
available HARPS\-pol and FORS\,2 spectropolarimetric observations obtained between 2014 and 2015.
Their analysis of line profiles and radial velocity shifts using 
HARPS\-pol and FORS\,2 subexposures with different time durations indicated the presence of spectral 
variability on short timescales, but the degree of this variability was rather low,
just at the intensity level of 0.2 per cent and 1.7 per cent, depending on the integration time and the considered element. 

\begin{figure*}
 \centering 
        \includegraphics[width=0.32\textwidth]{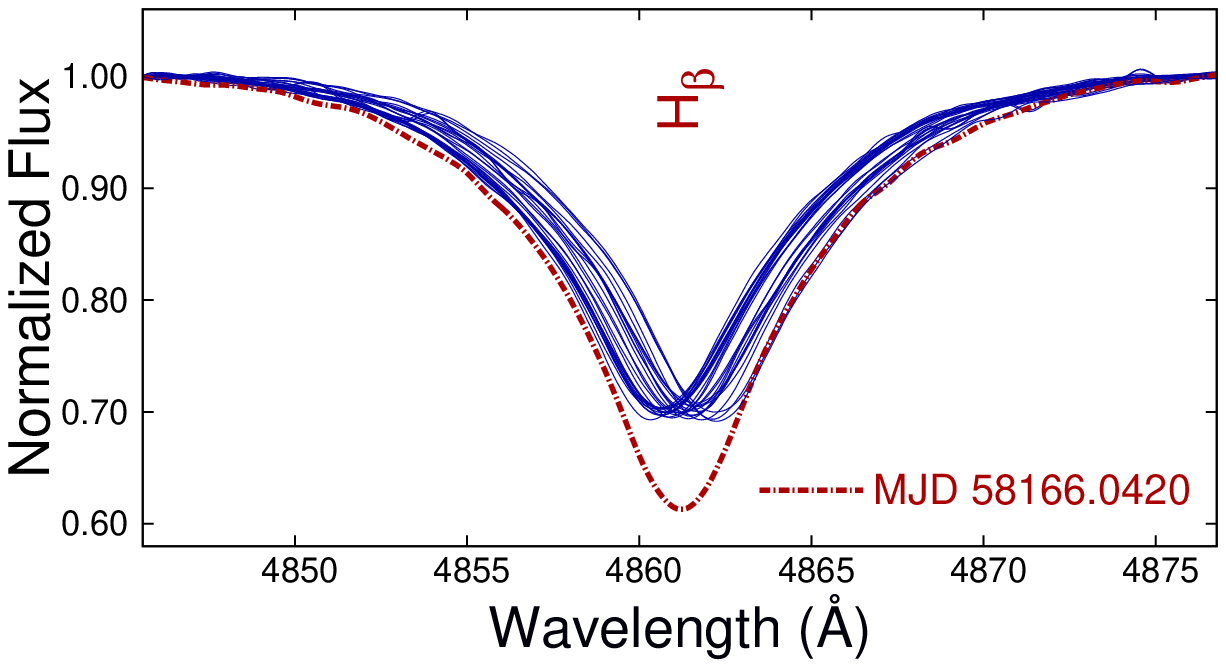}
        \includegraphics[width=0.32\textwidth]{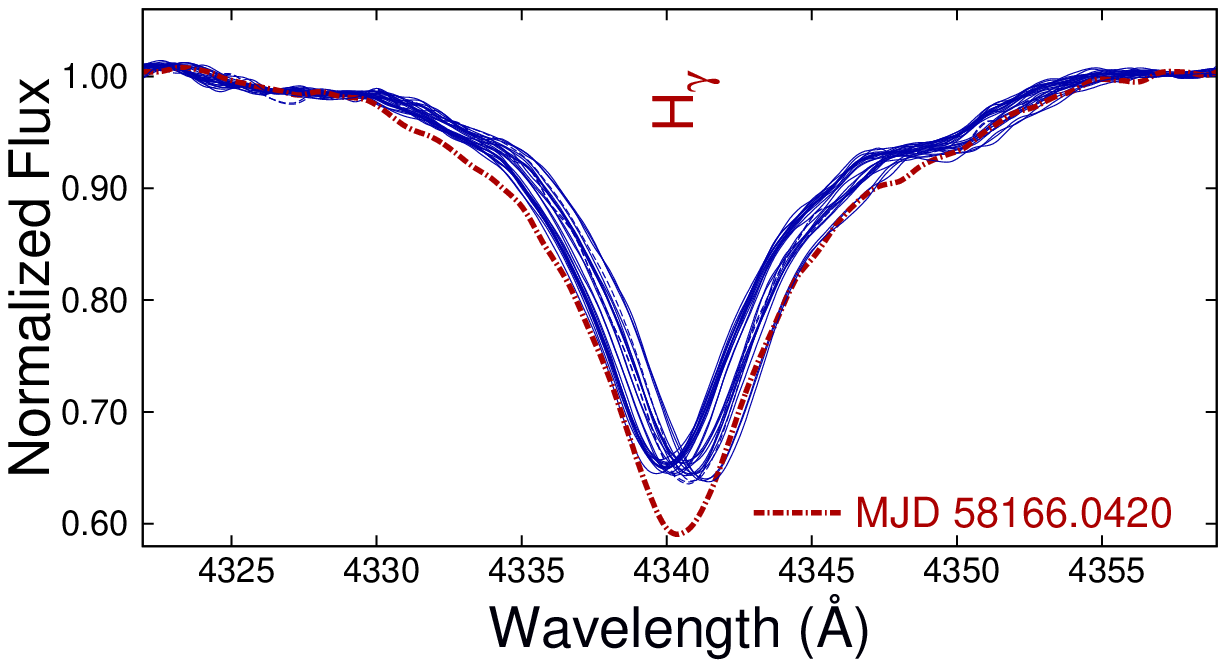}
        \includegraphics[width=0.32\textwidth]{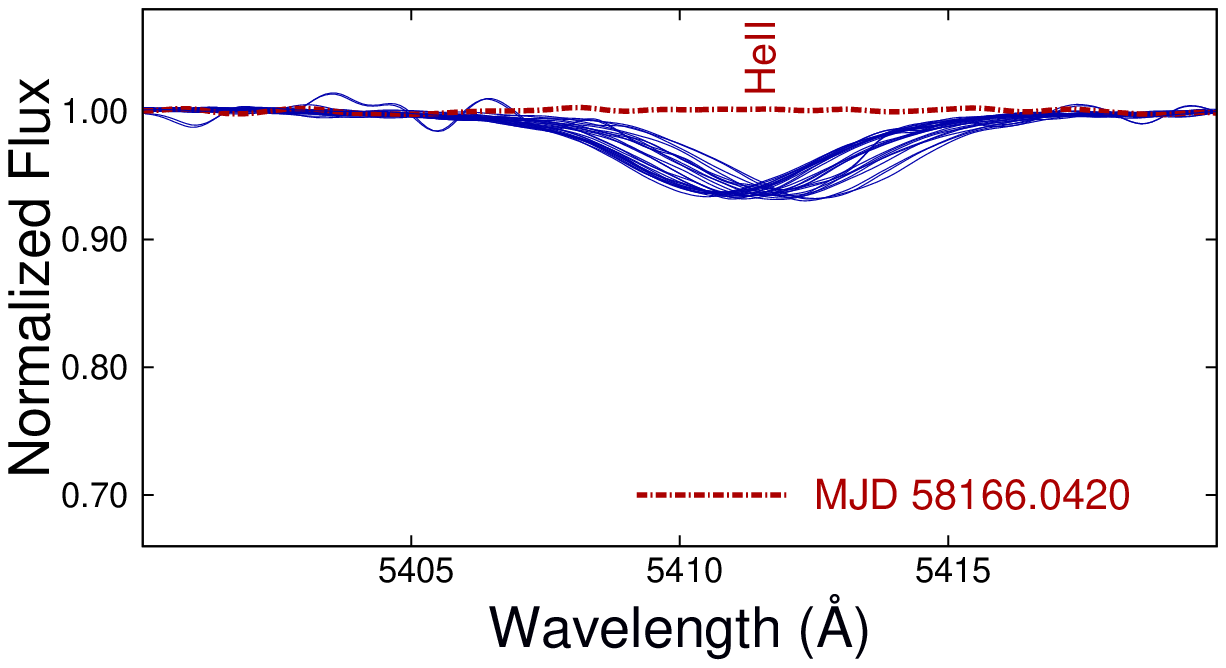}
        \includegraphics[width=0.32\textwidth]{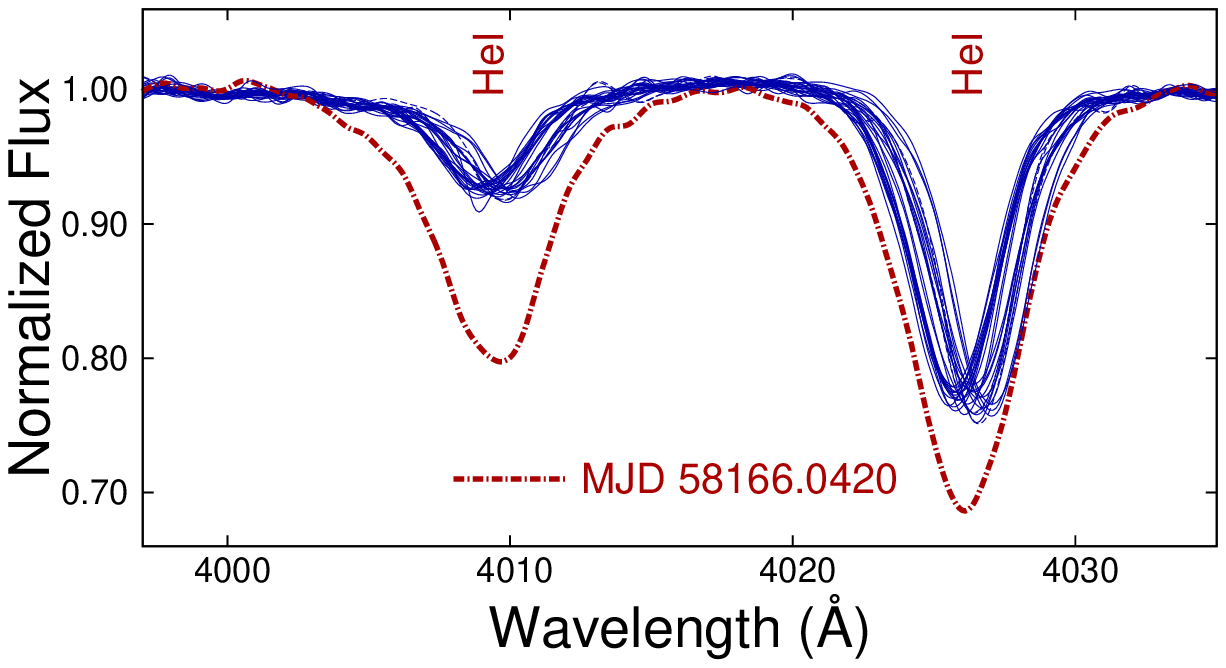}
        \includegraphics[width=0.32\textwidth]{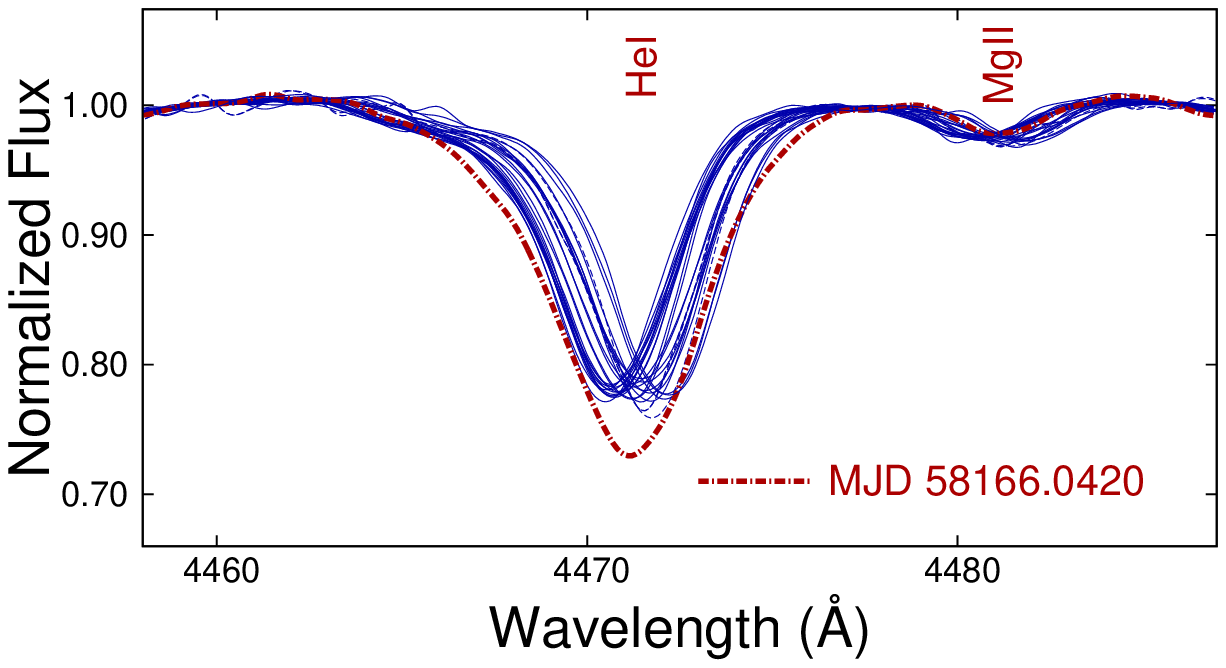}
        \includegraphics[width=0.32\textwidth]{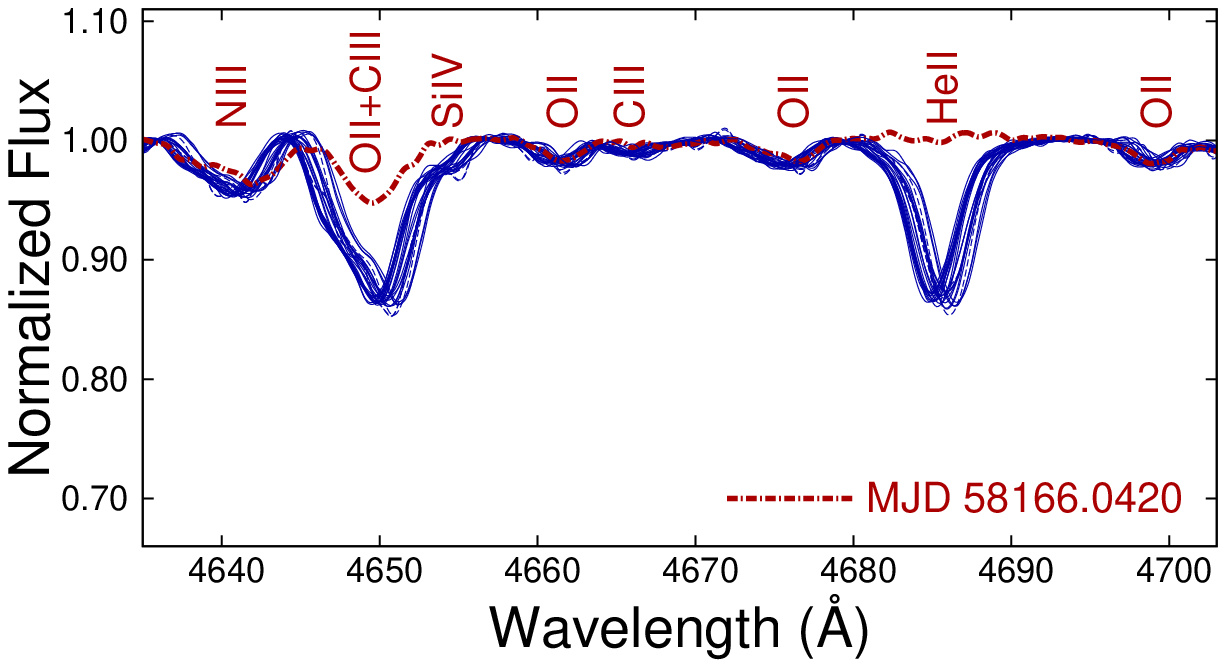}
        \caption{
Profile variability of hydrogen, \ion{He}{ii},  \ion{He}{i}, and metal lines.
         }
   \label{fig:varhydr}
\end{figure*}

\begin{figure}
 \centering 
        \includegraphics[width=0.95\columnwidth]{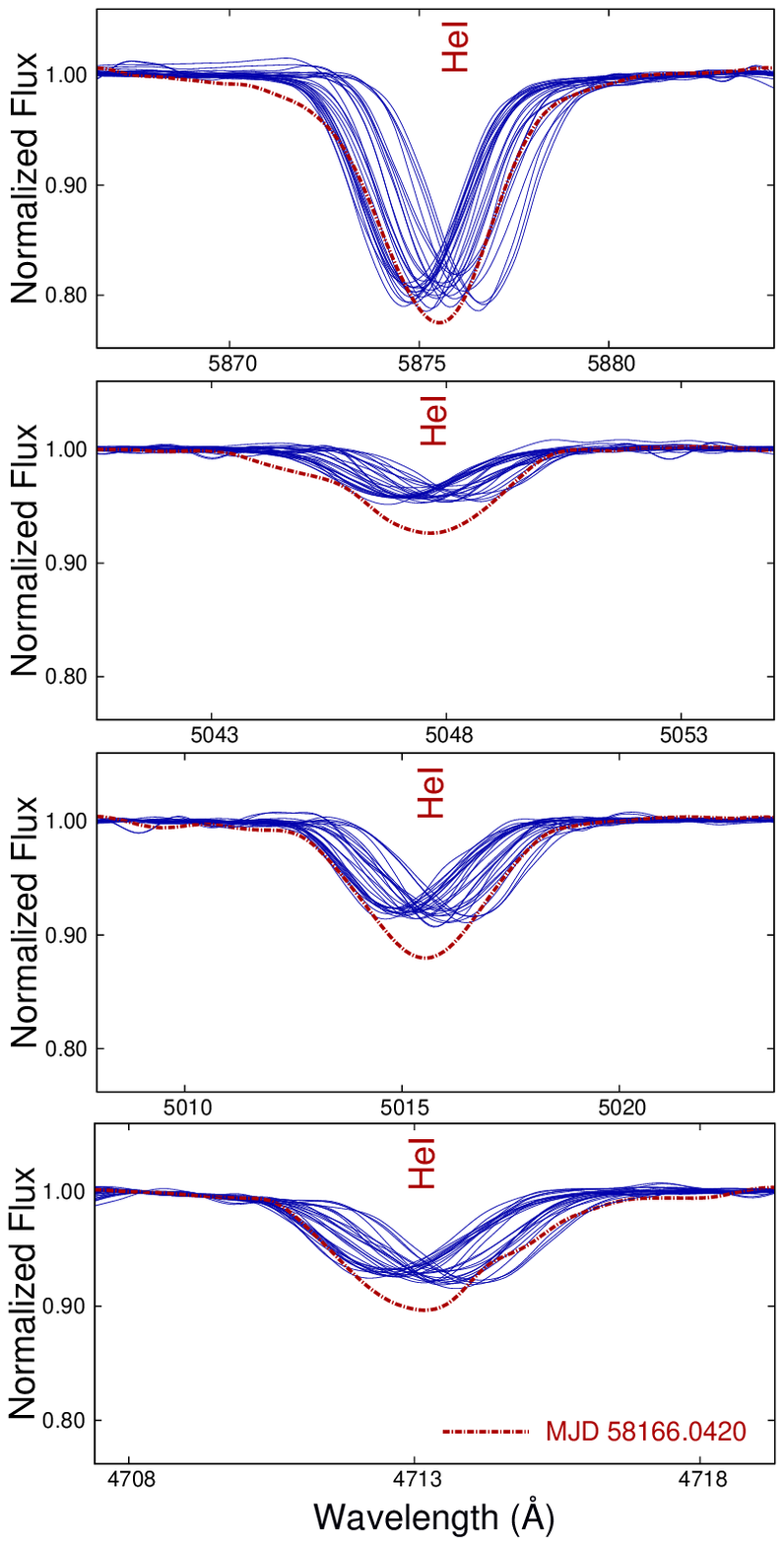}
        \caption{
Profile variability of a sample of  \ion{He}{i} lines.
         }
   \label{fig:varhe}
\end{figure}

In Figs.~\ref{fig:varhydr} and \ref{fig:varhe} we present several overplotted spectral lines 
belonging to different elements observed in FORS\,2 spectra acquired from 2014 to 2018. Evidently, 
all line profiles 
show distinct radial velocity shifts and very similar profile shapes apart from the line profiles 
observed at MJD58166.0420 (indicated 
by the red line) during the sudden increase of the longitudinal magnetic field.
At this observing epoch, all absorption hydrogen and \ion{He}{i} lines become stronger, whereas 
higher ionisation lines like \ion{He}{ii}, \ion{C}{iii} and \ion{Si}{iv} turn out to be much weaker, or fully 
disappear. Also the wind sensitive \ion{He}{ii}\,4686 line becomes extremely weak. 
Such a spectral appearance indicates a cooler photospheric temperature at the time of 
the magnetic field increase.  

\begin{figure}
 \centering 
        \includegraphics[angle=0,width=0.95\columnwidth]{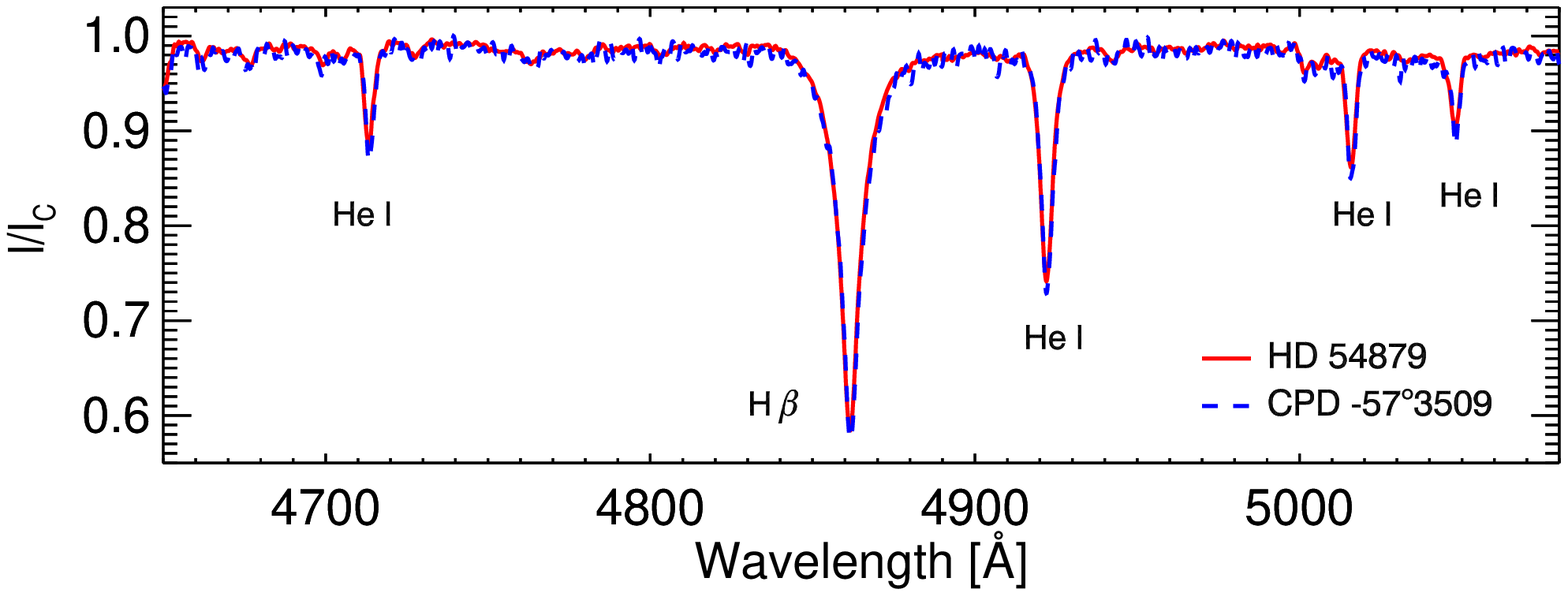}
        \caption{
Overplotted \ion{He}{i} and hydrogen line profiles of HD\,54879 observed on 2018 February 17
and the He-rich star CPD\,$-$57$^{\circ}$\,3509 of spectral type B1V.
         }
   \label{fig:cpd57}
\end{figure}

To estimate the decrease of the photospheric temperature, we compared the spectrum of HD\,54879 acquired on 
2018 February 17 with a few other early-B type stars well-studied with FORS\,2 in the framework of the
ESO large program ``B-fields in OB stars''. In Fig.~\ref{fig:cpd57}, we present overplotted 
\ion{He}{i} and hydrogen line profiles of HD\,54879 and the He-rich star  CPD\,$-$57$^{\circ}$\,3509
of spectral type B1\,V. The atmospheric parameters of CPD\,$-$57$^{\circ}$\,3509, 
$T_{\rm eff}=23750\,\pm\,250$\,K and $\log\,g= 4.05\,\pm\,0.10$, were obtained 
by \citet{Przybilla2016} using a high-resolution high signal-to-noise HARPS\-pol
spectrum.
The similarity of both stellar spectra suggests that the spectral type of HD\,54879 changed from O9.7\,V to  B1\,V
during the night when we observe the sudden magnetic field increase. It is very uncommon that spectral type changes 
in different rotation phases are observed in massive stars.
To our knowledge, only $a$\,Cen, a massive star with an anomalous He surface distribution, was
reported as a helium-weak star at one extremum of the magnetic field and as helium-rich at the
other, indicating B3 and B8 spectral types, respectively \citep{Norris1968}. However, in that star, the Balmer lines 
were invariant throughout the He line variation. This is not the case presented here for HD\,54879.

\begin{figure}
 \centering 
        \includegraphics[width=0.95\columnwidth]{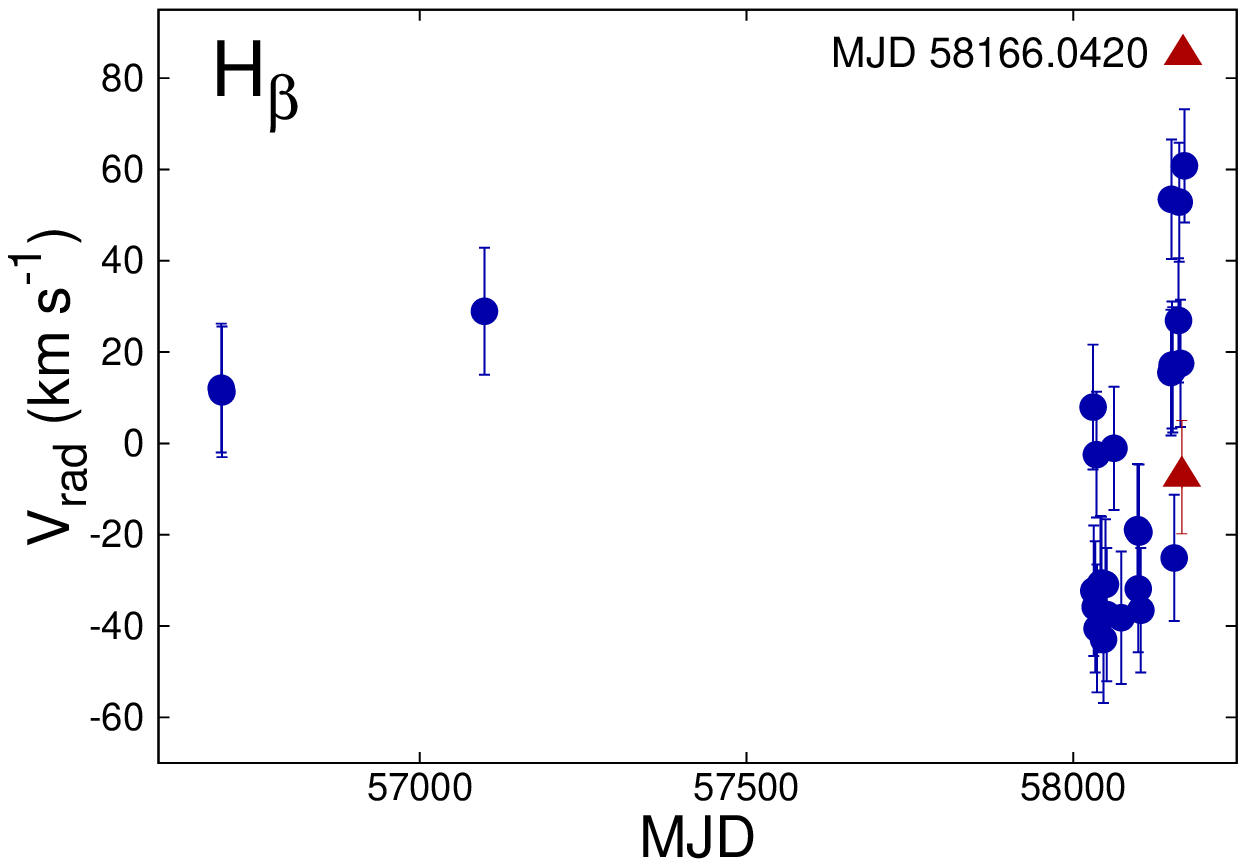}
        \includegraphics[width=0.95\columnwidth]{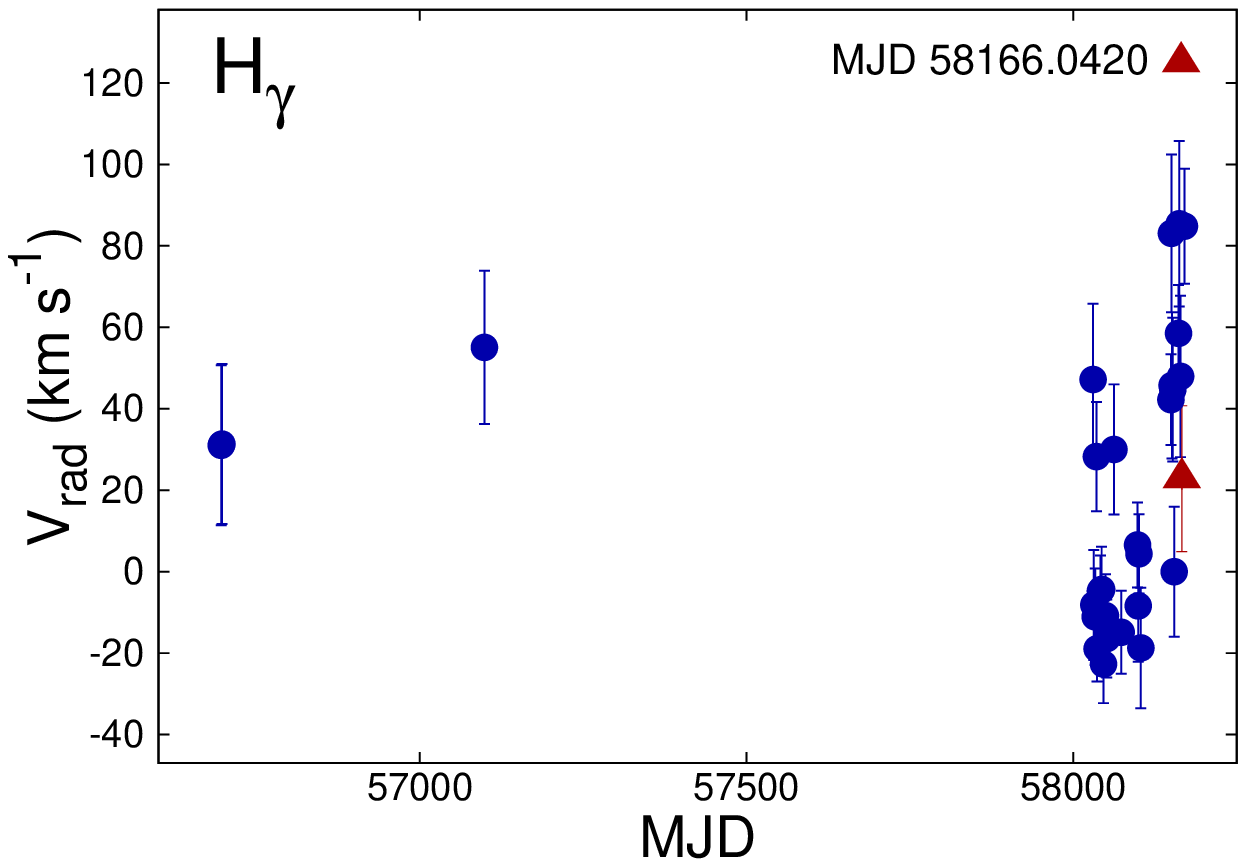}
        \includegraphics[width=0.95\columnwidth]{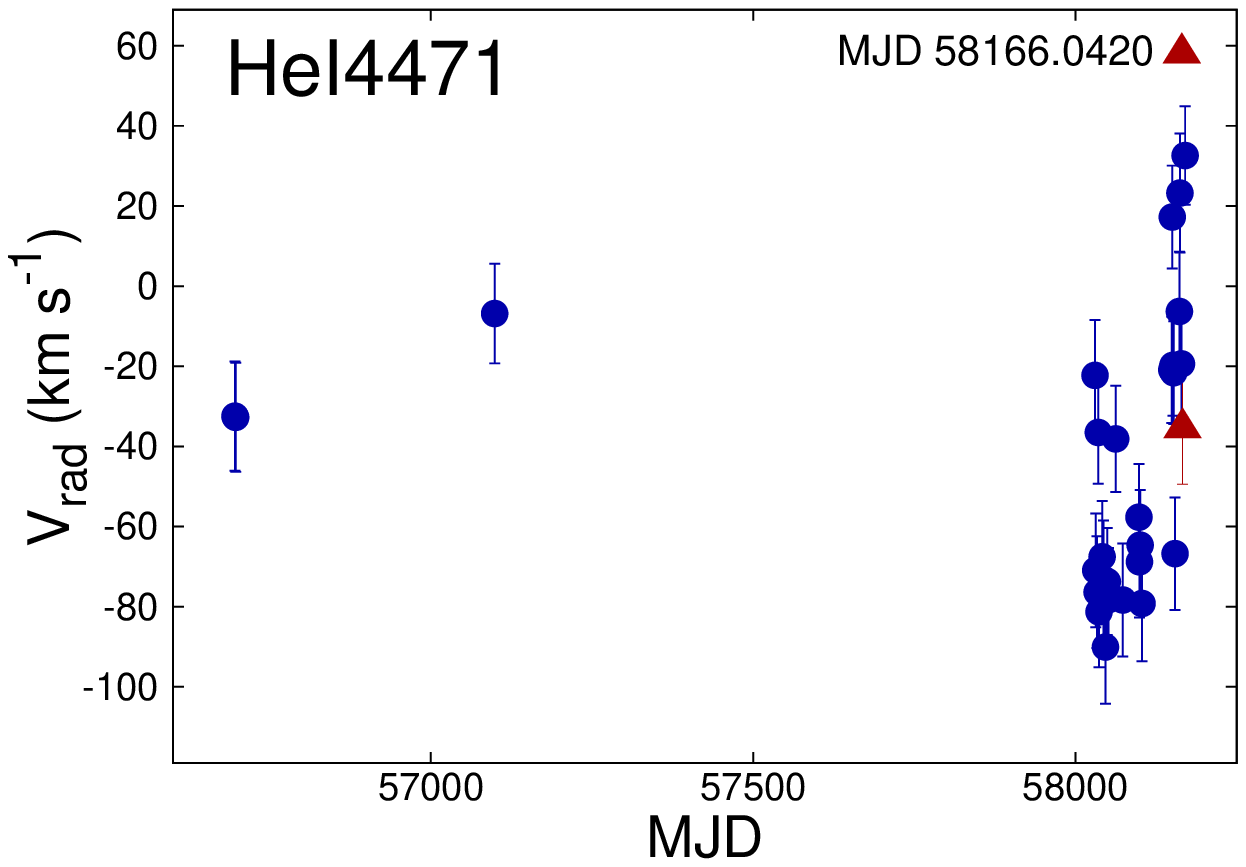}
        \includegraphics[width=0.95\columnwidth]{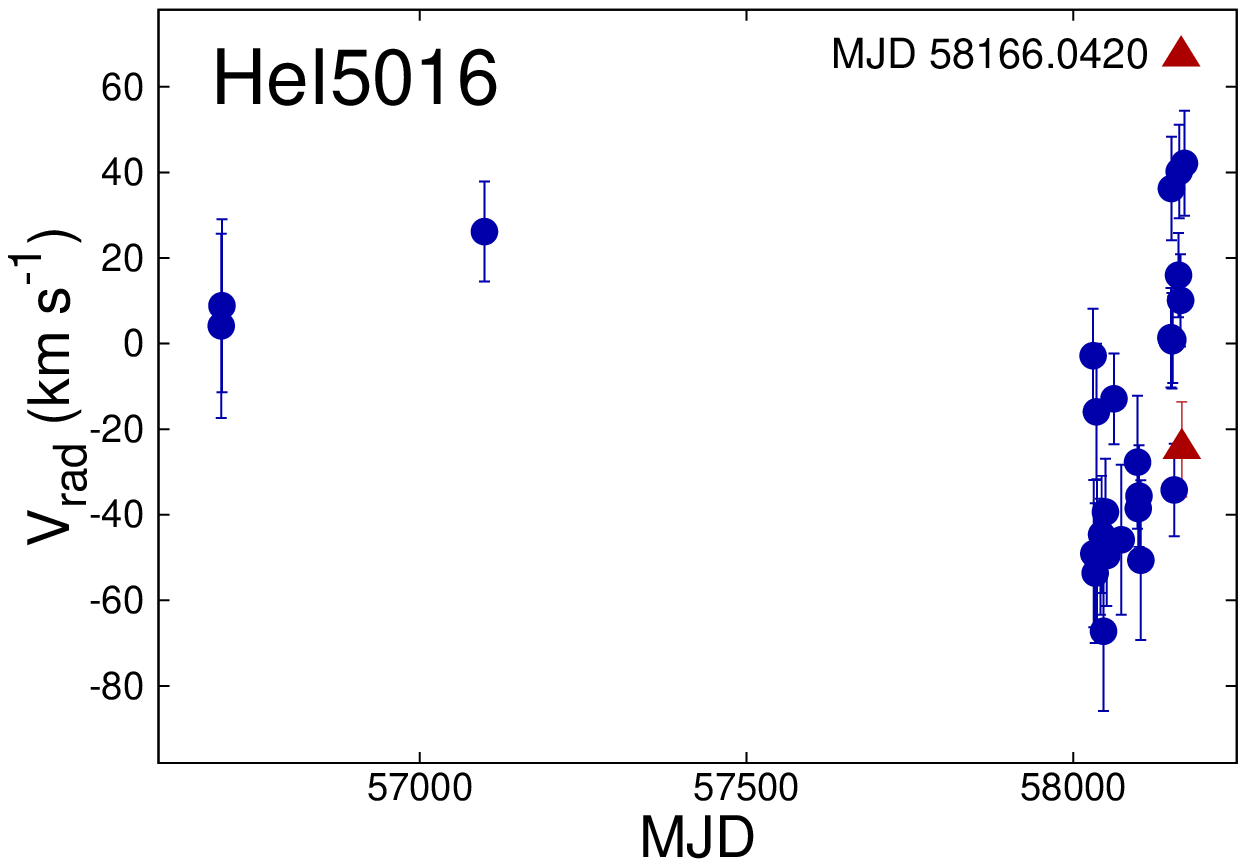}
                \caption{
Radial velocities measured for different spectral lines in the observations obtained between 2014 and 2018.
         }
   \label{fig:RV1evol}
\end{figure}

\begin{figure}
 \centering 
        \includegraphics[width=0.95\columnwidth]{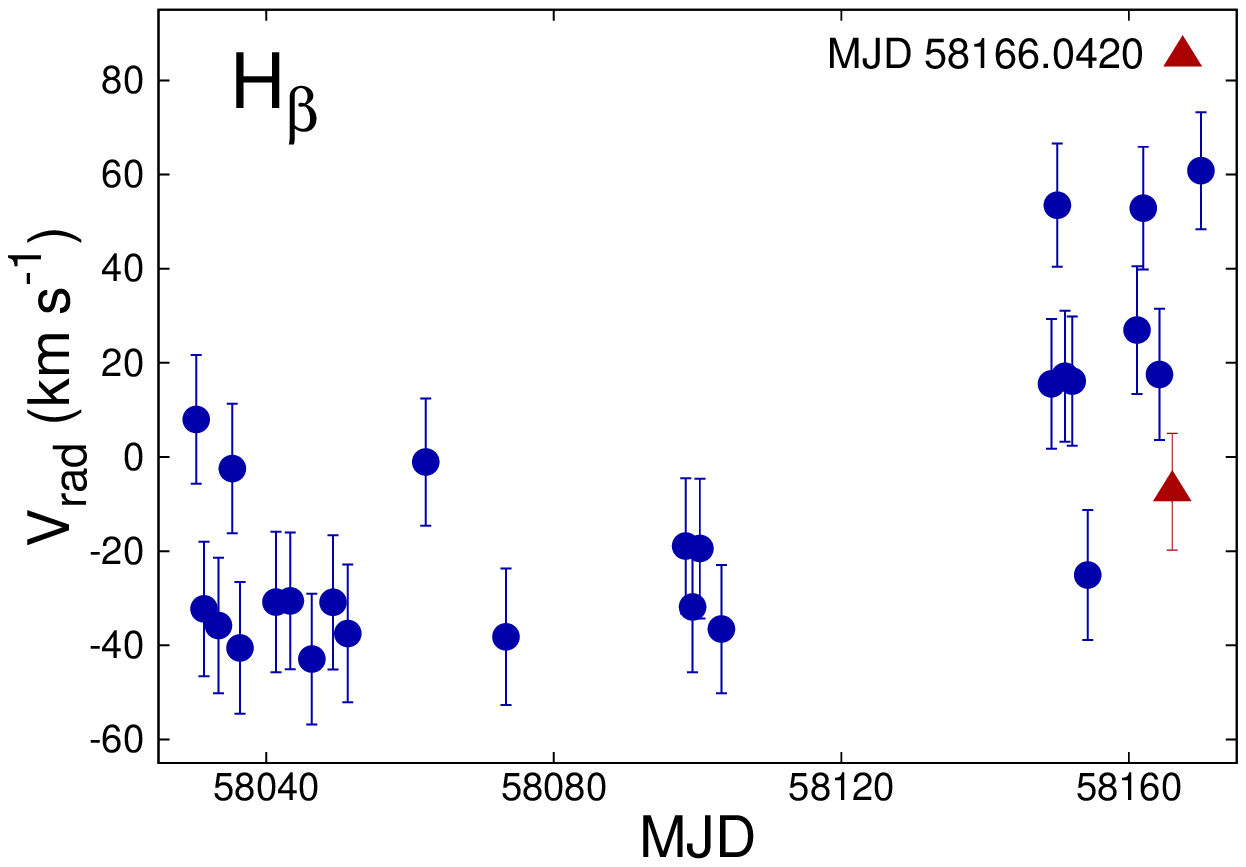}
        \includegraphics[width=0.95\columnwidth]{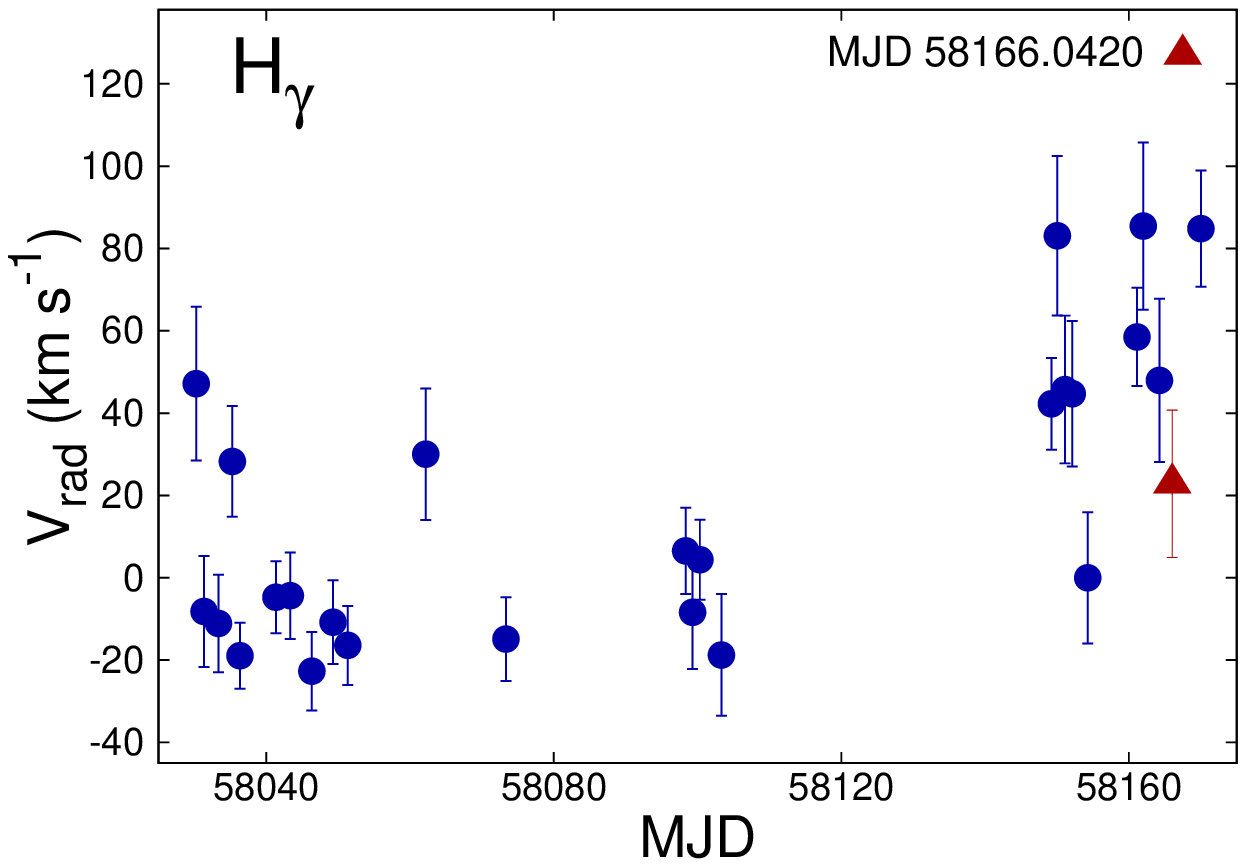}
        \includegraphics[width=0.95\columnwidth]{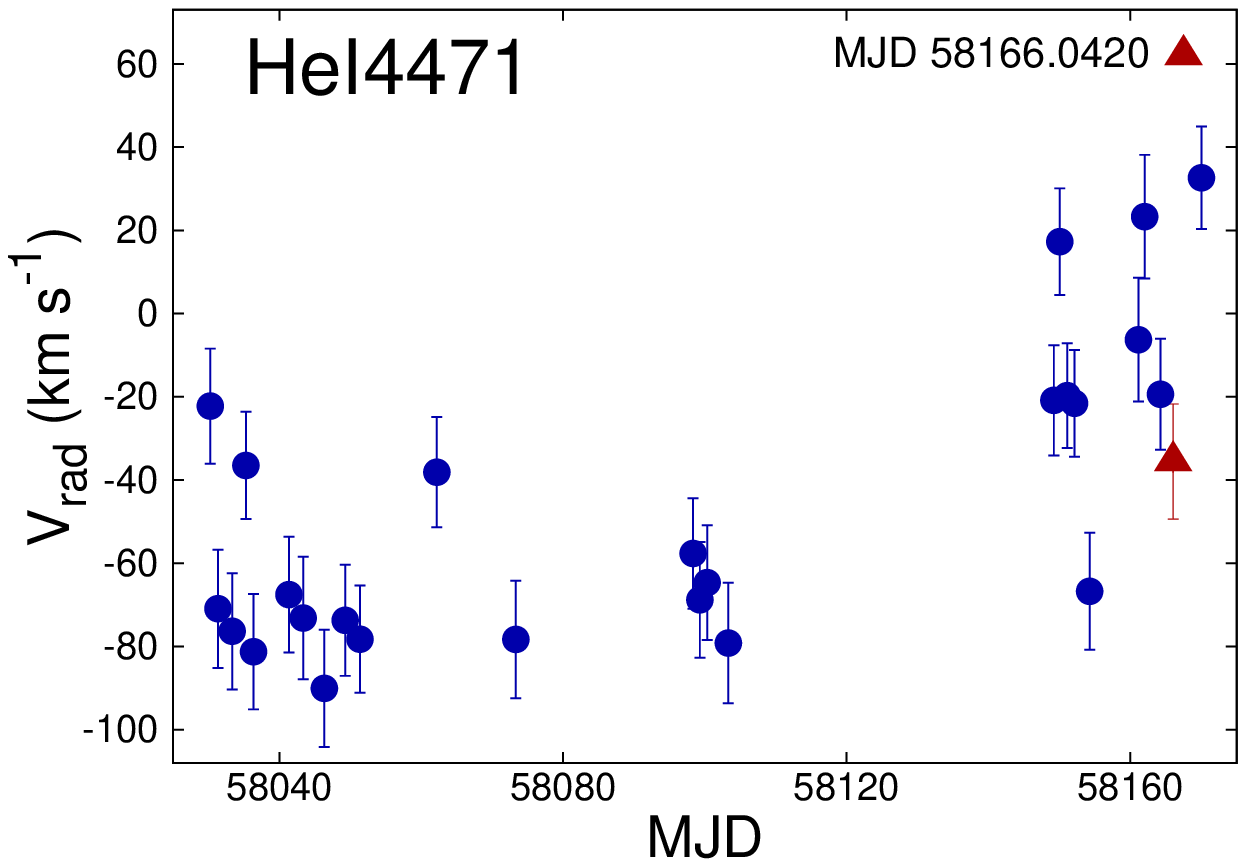}
        \includegraphics[width=0.95\columnwidth]{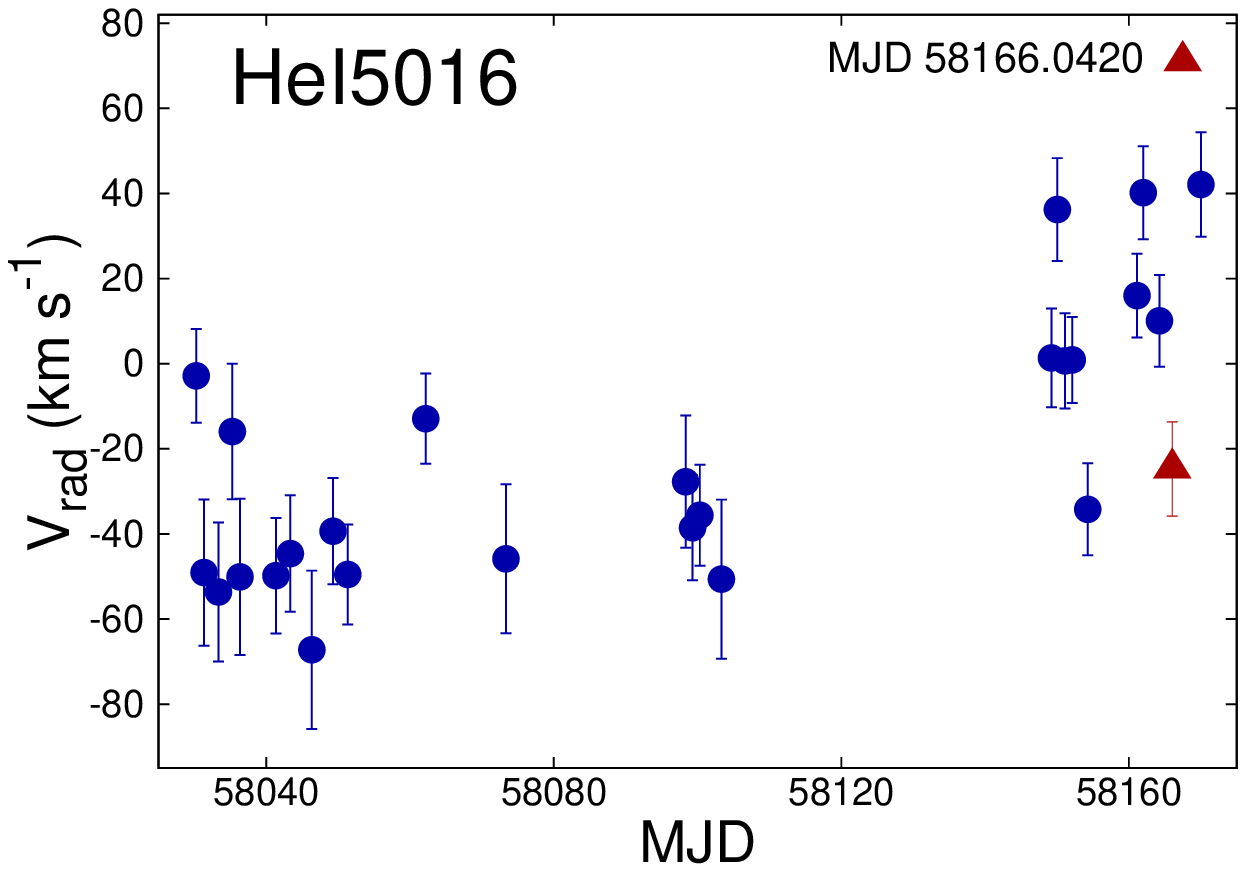}
                \caption{
Radial velocities measured for different spectral lines in the observations obtained between 2017 and 2018.
         }
   \label{fig:RV2}
\end{figure}

\begin{figure*}
 \centering 
        \includegraphics[width=0.32\textwidth]{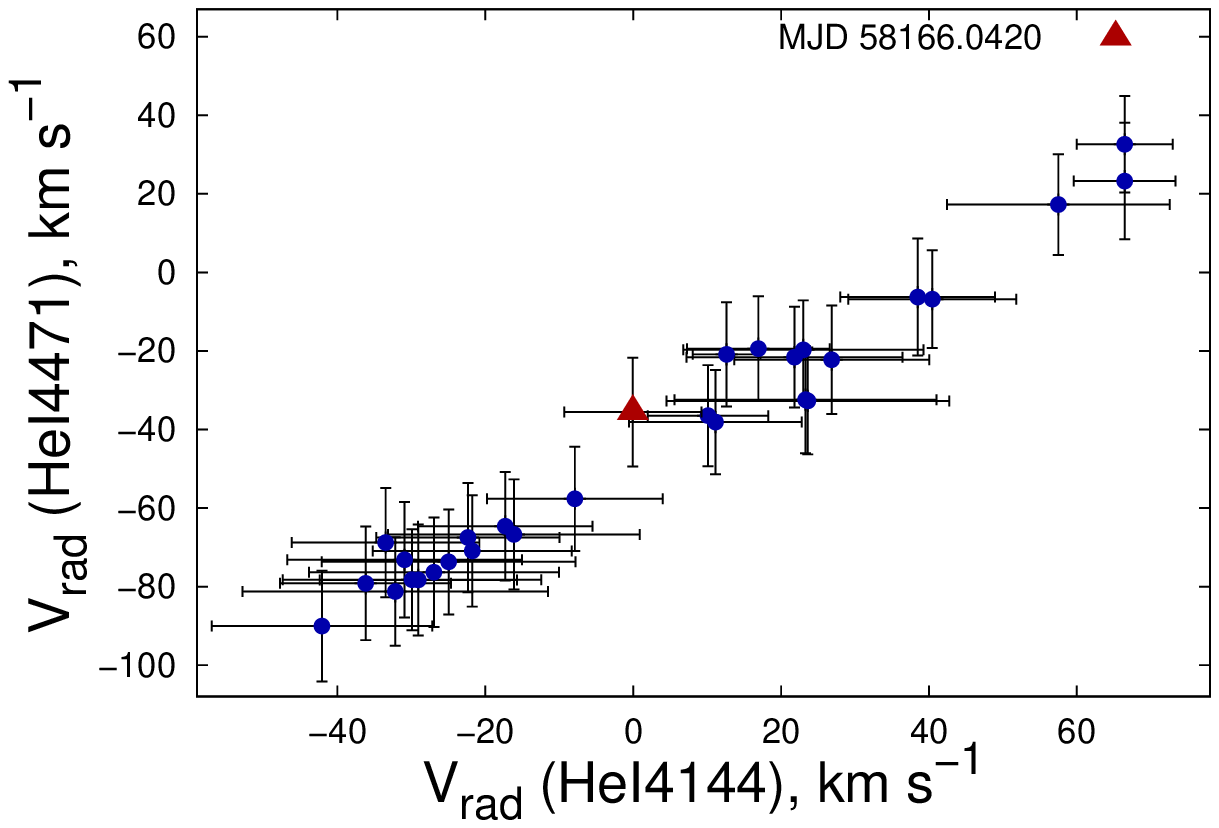}
        \includegraphics[width=0.32\textwidth]{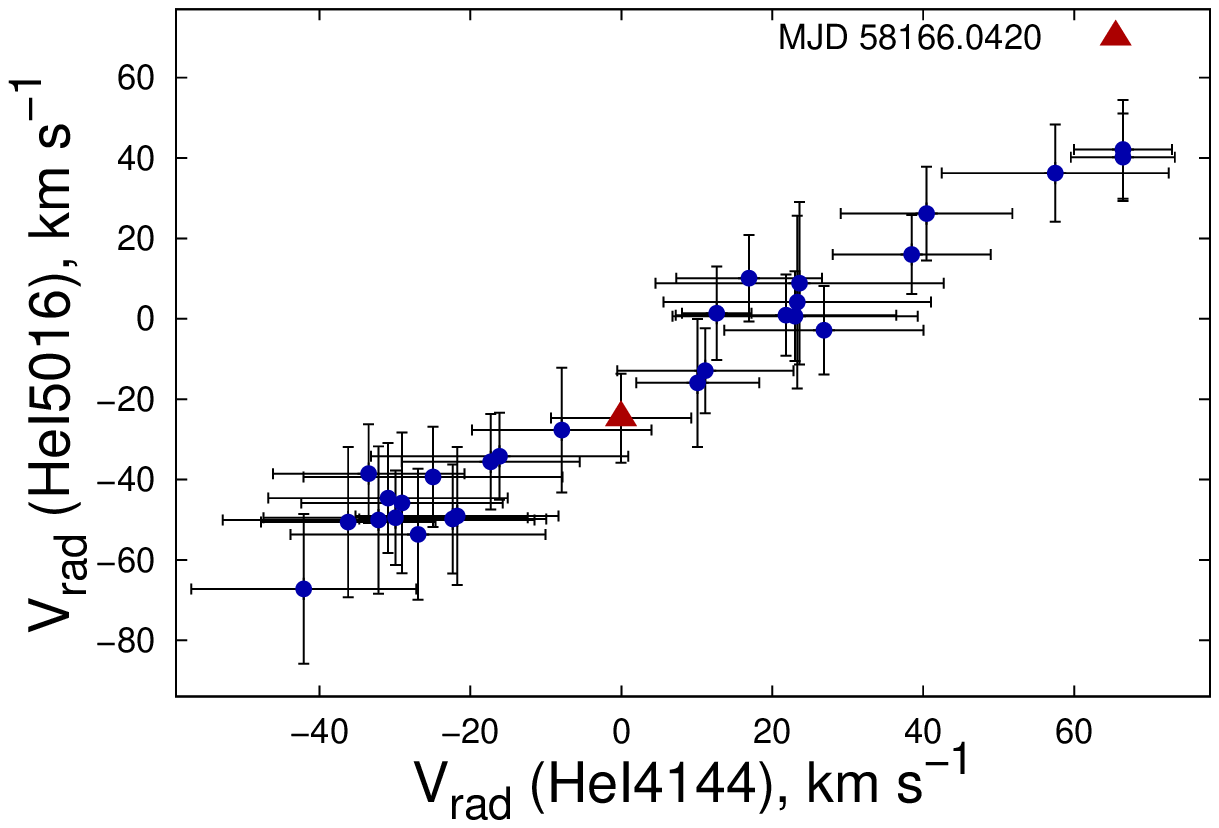}
        \includegraphics[width=0.32\textwidth]{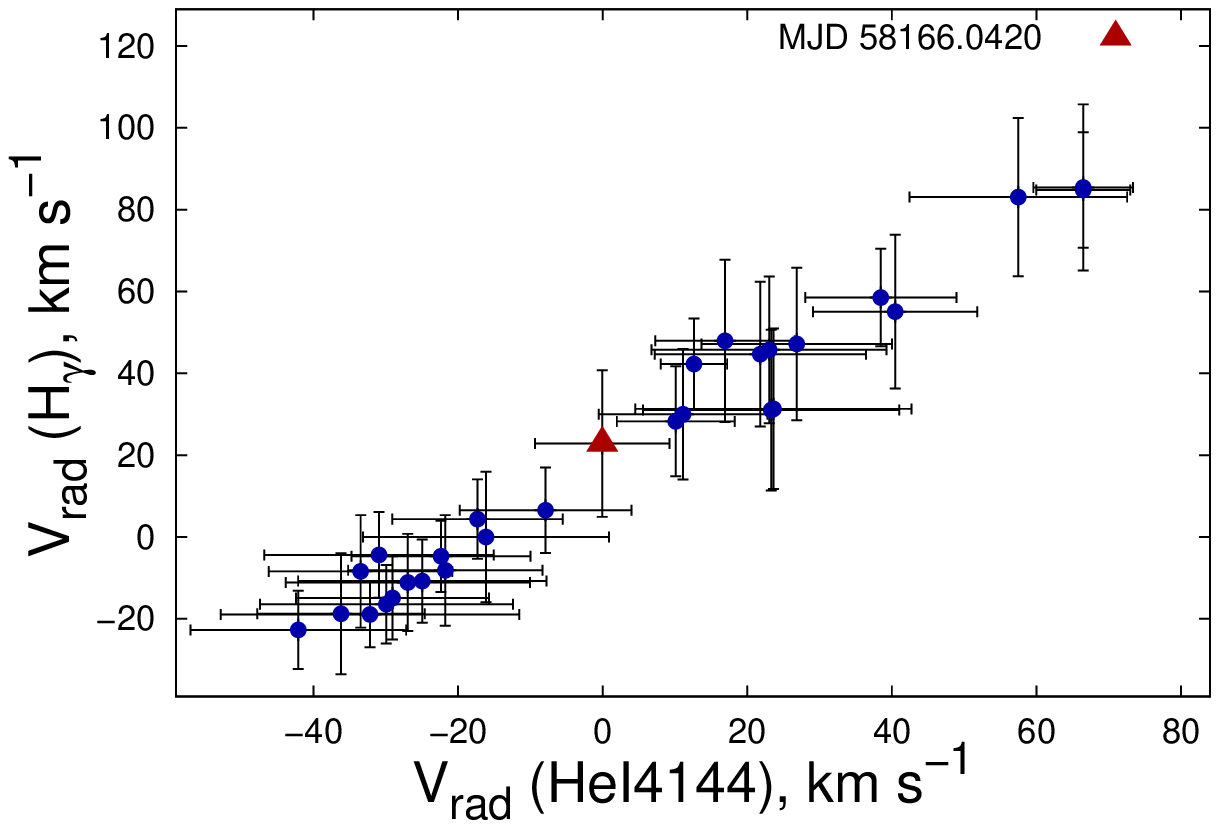}
        \includegraphics[width=0.32\textwidth]{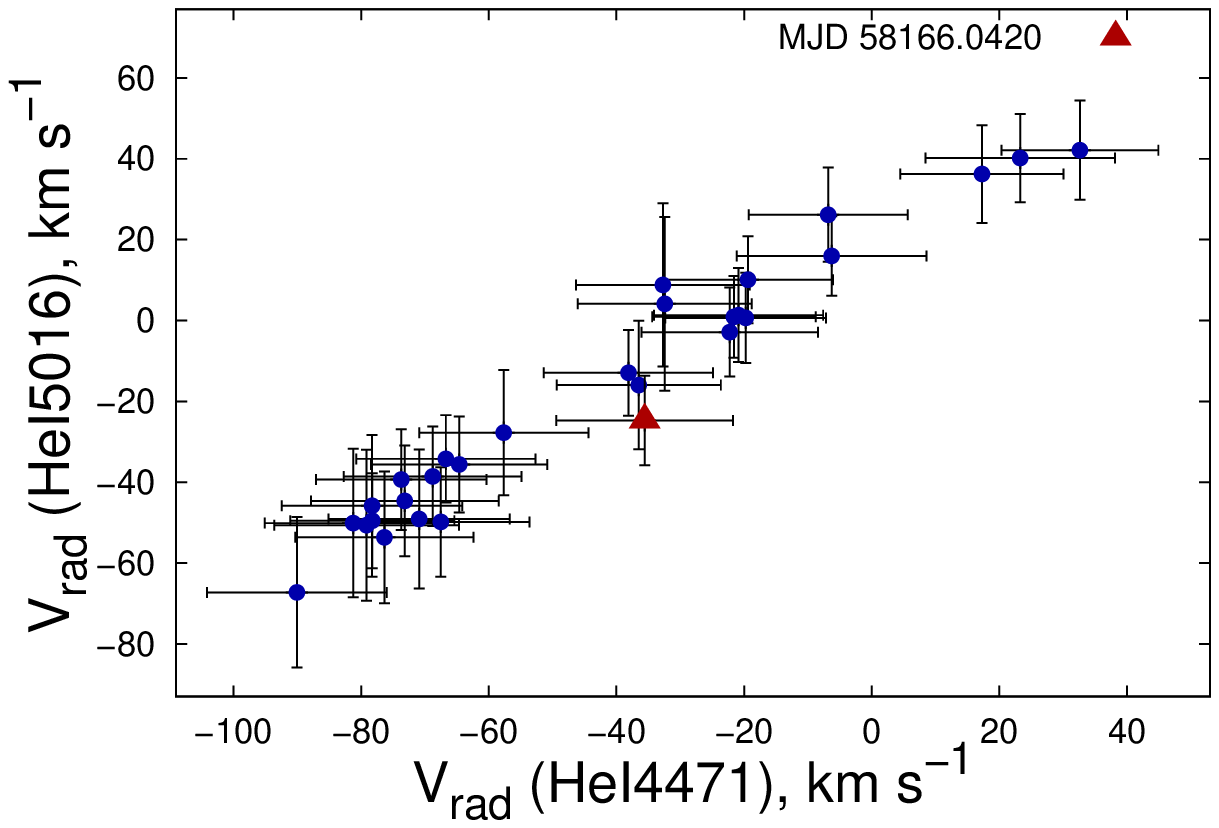}
        \includegraphics[width=0.32\textwidth]{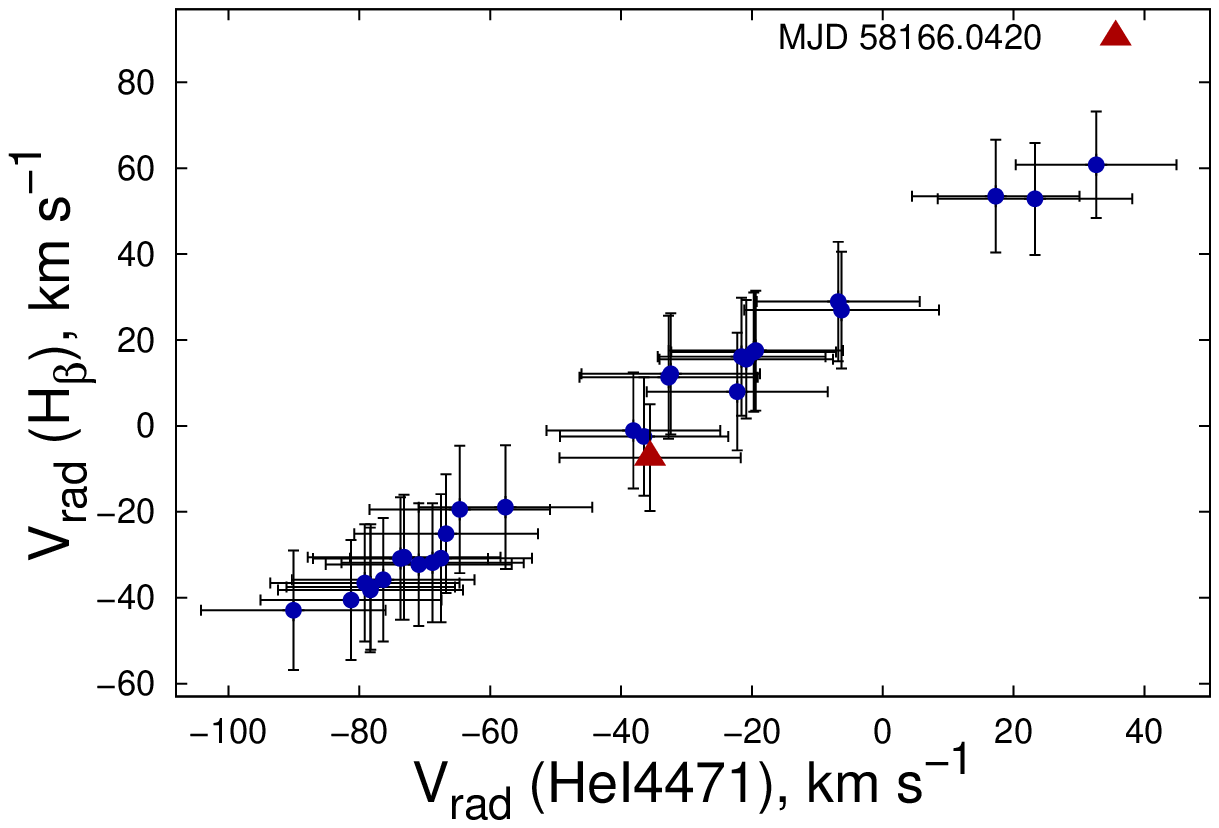}
        \includegraphics[width=0.32\textwidth]{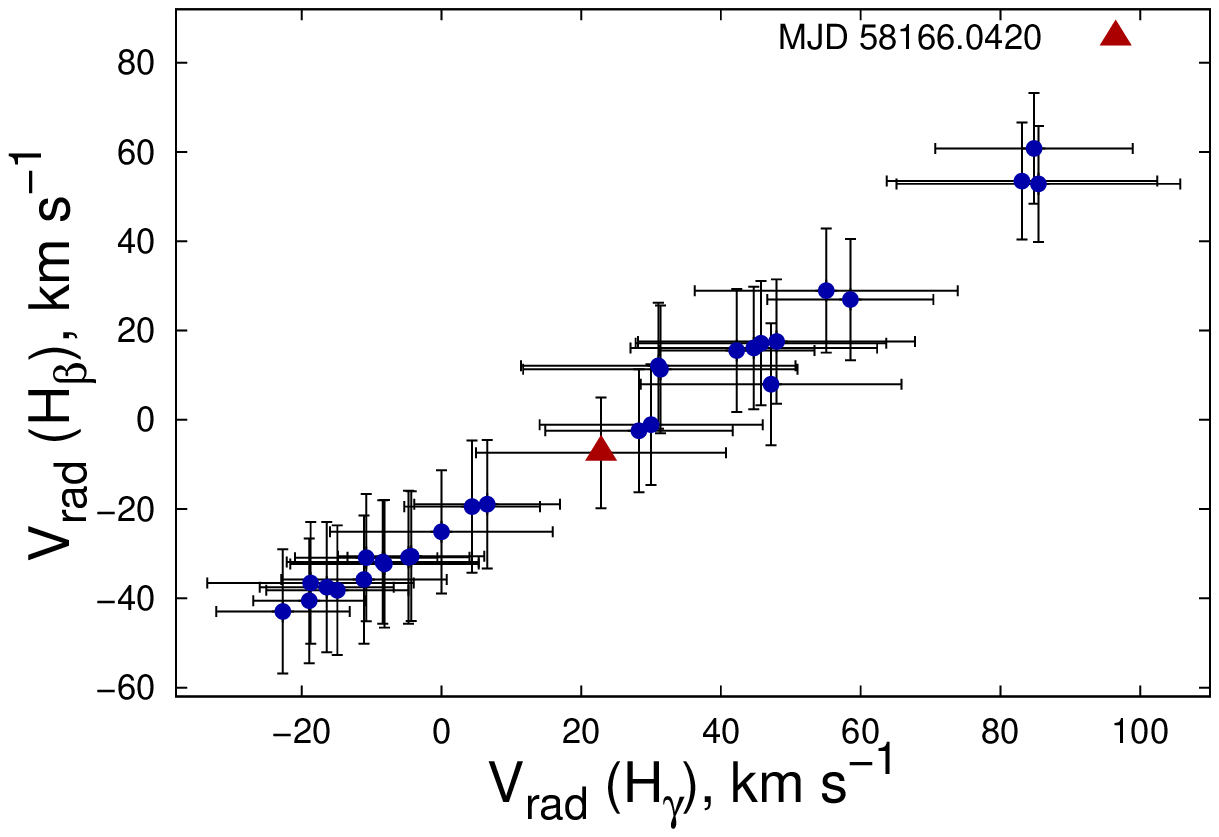}
        \caption{
Radial velocity correlations for hydrogen and \ion{He}{i} lines.
         }
   \label{fig:rvcor}
\end{figure*}

The radial velocity changes for different lines in all FORS\,2  spectra acquired between 
2014 and 2018 are presented in Fig.~\ref{fig:RV1evol} and those measured in our recent 
spectra between 2017 and 2018 in Fig.~\ref{fig:RV2}. While not much can be concluded about the variability
of the radial velocities in 2014--2015 due to the rather large measurement uncertainties, 
the velocity changes in 2017--2018 are more pronounced, indicating an increase by $\sim100$\,km\,s$^{-1}$.
Remarkably, we observed at the epoch of the sudden magnetic field
increase (indicated by the red triangle) that the radial velocity of all spectral lines 
dropped by several 10\,km\,s$^{-1}$.
To better highlight the observed changes in radial velocities, we show in Fig.~\ref{fig:rvcor}
the radial velocity shift correlations between the lines belonging to different elements. 

The observed dispersion of the radial velocity measurements in these figures is probably related to
$\beta$~Cephei--like pulsations. The atmospheric parameters presented by
\citet{Castro2015} suggest that HD\,54879 has already
slightly evolved from the ZAMS and is passing through the $\beta$~Cephei
instability strip. As mentioned above, \citet{Jarvinen2017} were not able to detect 
significant velocity shifts in high-resolution
  HARPS\-pol observations of HD\,54879, most likely due to much longer exposure
  times, of the order of 1--3\,h. During the long HARPS\-pol exposures, any spectral variability is smeared
over the pulsation cycle and difficult to detect. In contrast, FORS\,2 observations carried out
using an 8\,m telescope, have a duration of only 10--20\,min and are expected to be more strongly affected 
by the $\beta$~Cephei--like pulsations. According to \citet{Telting2006},
at least half of the late O to early B solar-neighbourhood stars that
are located in the $\beta$~Cephei instability strip are actually  pulsating in radial
and/or non-radial modes. 

With respect to the remarkable change of the spectral appearance of HD\,54879 during the sudden 
magnetic field increase, it is important to mention that a recent study of the most strongly 
magnetized O-type star NGC\,1624-2
using UV observations with HST/COS revealed similar dramatic variations in the resonance line profiles
between rotation phases when looking nearly at the magnetic pole
and those when looking nearly at the magnetic equator
(see Fig.~1 in \citealt{David2018}). Similar to our observations of HD\,54879, these line profiles 
show at these two rotation phases very different characteristics including the line intensity,  line shape and width, 
and radial velocity. It should however be noted that the variability of the UV resonance lines in massive stars is 
usually related to the stellar wind or to the magnetosphere, while for HD\,54879 we observe line profile changes in the 
photospheric lines.

\begin{figure}
 \centering 
        \includegraphics[width=0.95\columnwidth]{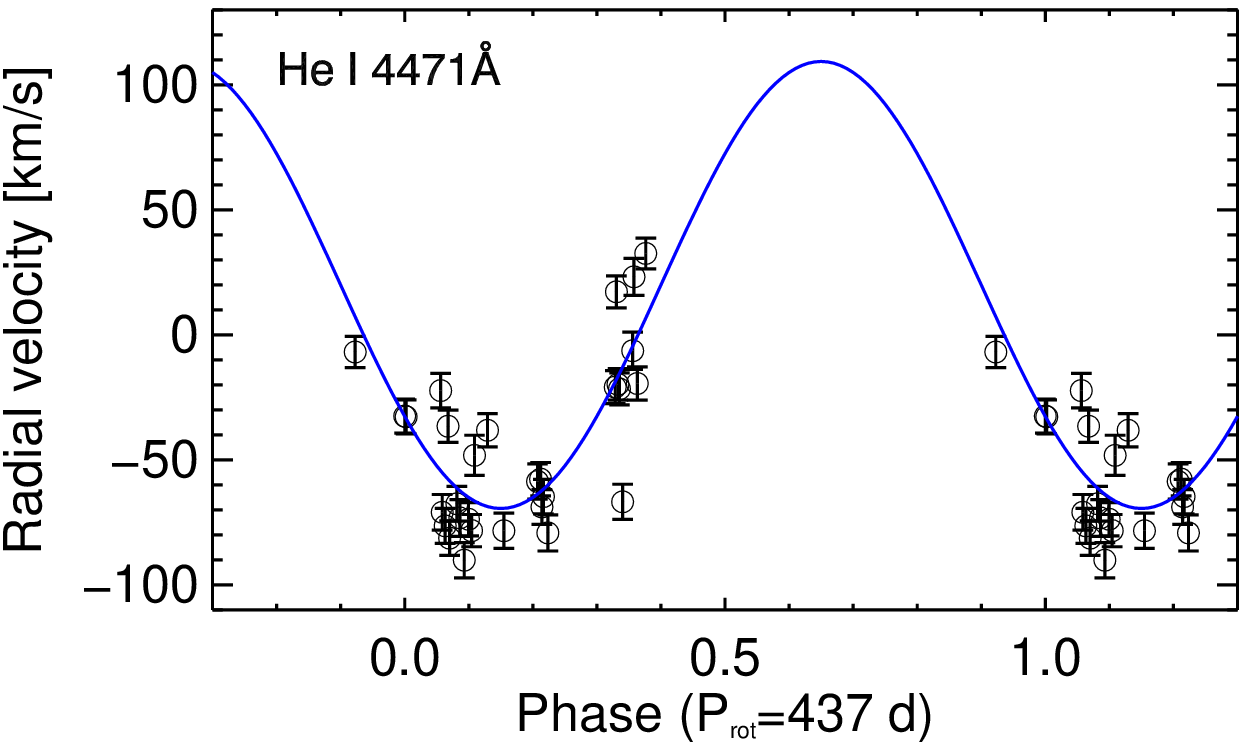}
\includegraphics[width=.95\columnwidth]{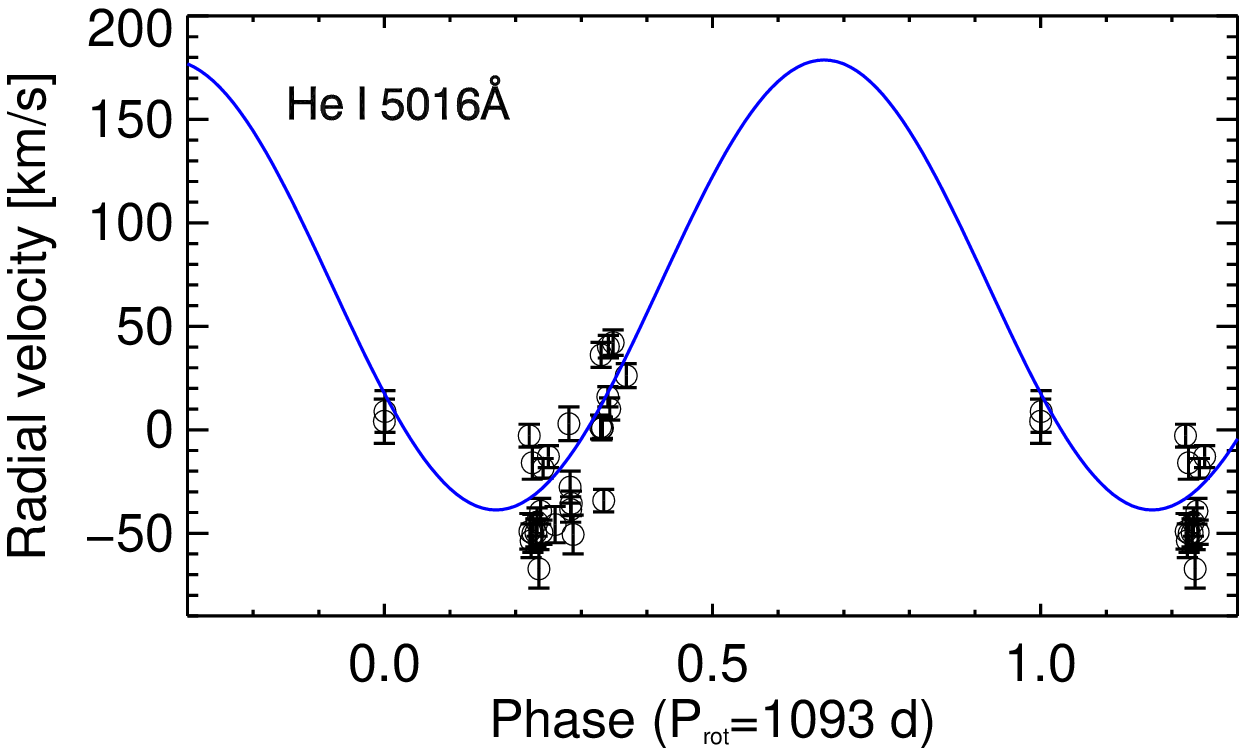}
                \caption{
{\it Upper panel:} Radial velocities of the \ion{He}{i}~4471 line measured in the spectra obtained 
between 2014 and 2018 are fitted with a period of 437\,d. {\it Lower panel:}
Radial velocities of the \ion{He}{i}~5016 line fitted with a period of 1093\,d.
         }
   \label{fig:per}
\end{figure}

It cannot be excluded that 
the observed changes in radial velocity are caused by an invisible wide companion.
Indeed, 
\citet{Castro2015} compared their radial velocity measurement from one HARPS\-pol spectrum
with other measurements 
in the literature and concluded that HD\,54879 could be a member of a long-period binary system.
Our search for periodicity in radial velocities was however rather inconclusive:
using radial velocity changes in the line \ion{He}{i}~5016 we find an indication of a possible period of 
1093\,d, while the data for the  \ion{He}{i}~4471 are better fitted with a period
of 437\,d. To search for the period, we used 
a  non-linear  least-squares  fit  to  multiple  harmonics  using  the
Levenberg-Marquardt method \citep{Press1992}. To detect the
most probable period, we calculated the frequency spectrum, and
for each trial frequency we performed a statistical F-test of the
null hypothesis for the absence of periodicity \citep{Seber1977}. The
resulting F-statistics can be thought of as the total sum, including  covariances  of  
the  ratio  of  harmonic  amplitudes  to  their standard deviations, i.e.\ an S/N.
The measured radial velocities for both lines fitted with the corresponding periods are presented in 
Fig.~\ref{fig:per}.
Obviously, due to the poor spectral coverage in the years between 2014 and 2017 and the observed large 
dispersion of the radial velocity measurements, our results from the period search are only tentative and
should be verified by additional spectropolarimetric observations.

\section{Discussion}

It is the first time that a sudden increase of the magnetic field strength at a rotation phase close to 
the magnetic equator is observed in a magnetic 
massive O-type star. As the majority of massive O-type stars rotate rather slowly and rarely are 
spectroscopically and polarimetrically  monitored on several consecutive nights, it is not possible 
to know whether other magnetic massive stars show sudden increases of their magnetic field strength. 
On the other hand, the remarkable  change of the 
spectrum of HD\,54879 during the magnetic field increase should be easily detected in highly accurate photometric observations.

The nature of magnetic fields in massive stars is still unknown. The presence of magnetic activity is expected 
since stellar spots, X-ray emission, and flares are observed in OB-type stars (e.g.\ \citealt{Groote2004, 
Mullan2009, Smith1993, Reiners2000}). Such activity indicators can be 
caused by the presence of arcs and filaments, which however are not discovered yet in massive stars.
Within such a scenario, the observed transition to a cooler spectral type and in particular the significant 
change of the radial velocity of the spectral lines at the time of the sudden magnetic field increase would indicate 
the presence of a giant temperature spot, which could be related to a large magnetic flux tube that breaks 
the surface of the photosphere. The dramatic change in radial velocity could then be related to
the flow velocity.
Similar phenomena are observed in the Sun and solar-like stars, but for smaller scale flux tubes.
We can only speculate that the observed cooler spectral type of HD\,54879 is probably caused
by the increased visibility of the deeper atmospheric layers in the temperature spot and the presence of a
non-standard temperature gradient in the upper layers of this massive star due to a large amount
of circumstellar absorption related to the large wind confinement radius.

\begin{figure*}
 \centering 
        \includegraphics[width=0.95\textwidth]{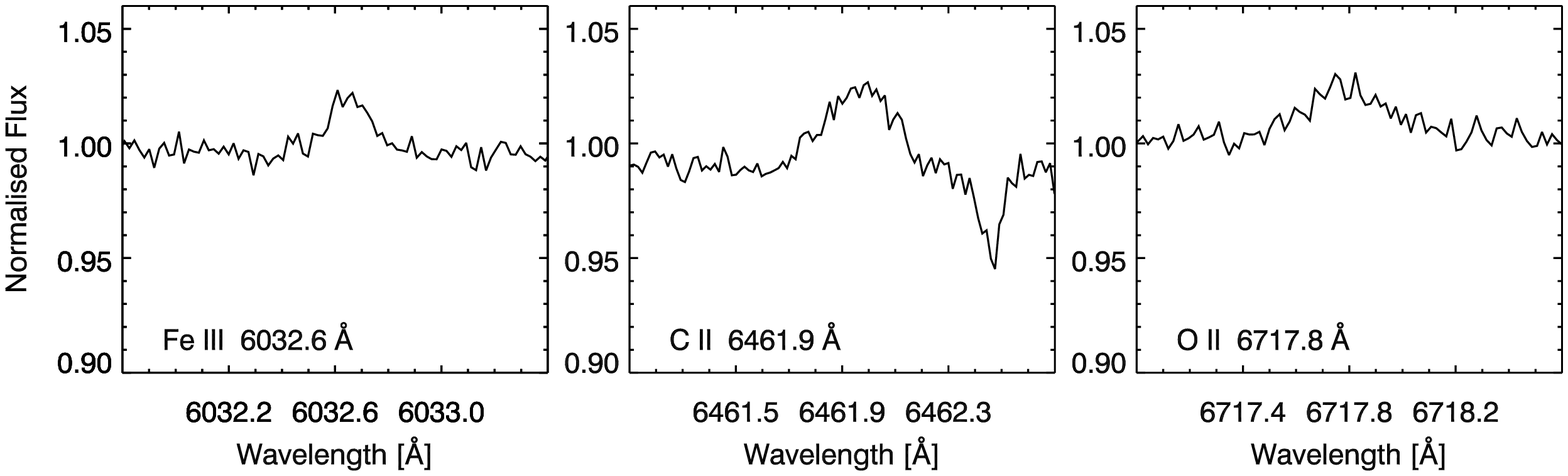}
                \caption{
Examples of emission line profiles in the high-resolution HARPS\-pol spectrum.
         }
   \label{fig:emis}
\end{figure*}

\begin{figure}
 \centering 
        \includegraphics[width=0.95\columnwidth]{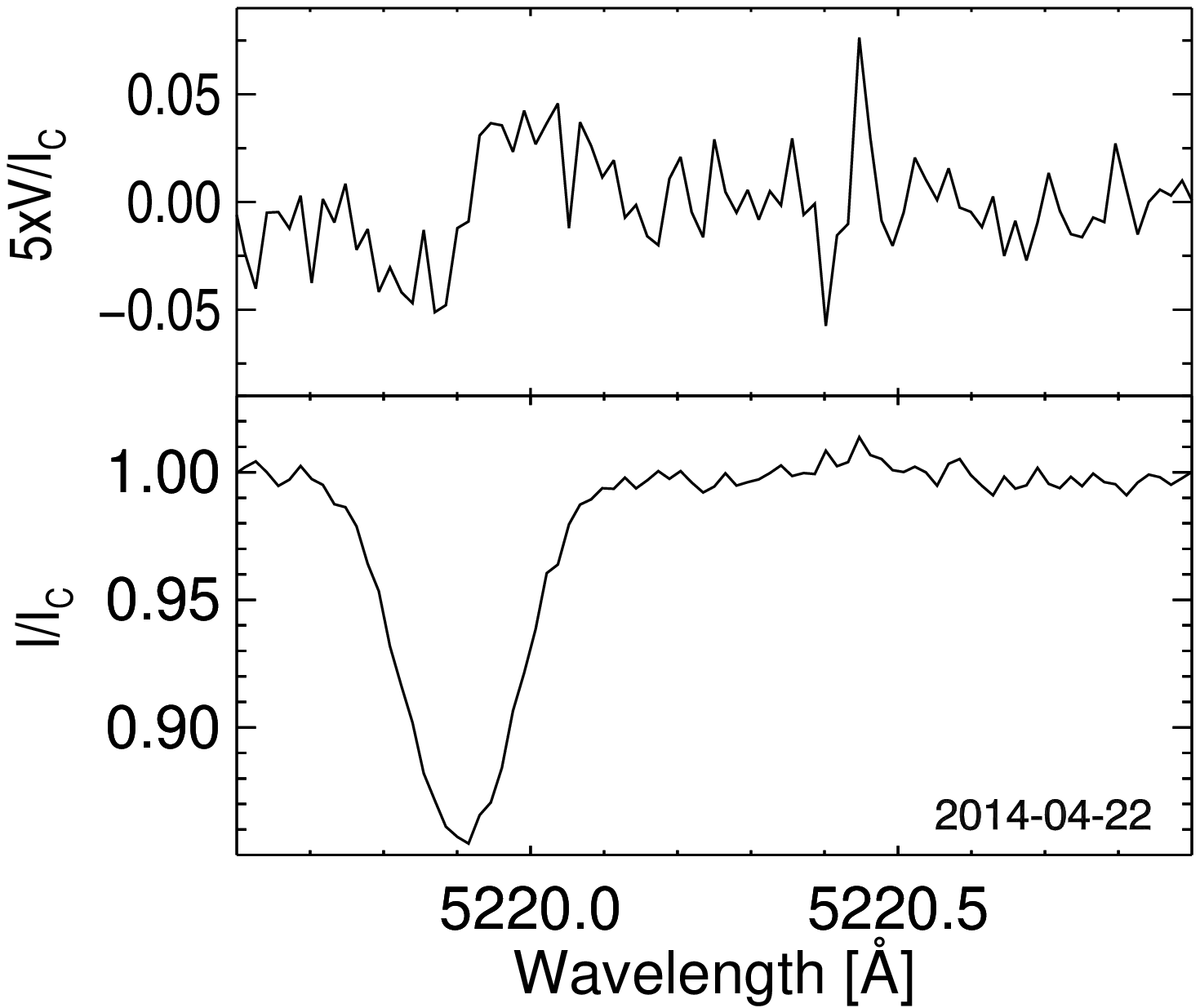}
                \caption{
Stokes~$I$ and Stokes~$V$ spectra in the vicinity of a Zeeman feature related to an unidentified emission 
line with an unknown Land\'e factor.
         }
   \label{fig:weak}
\end{figure}

Notably, our high-resolution HARPS\-pol spectra obtained within the framework of the 
ESO Large Prg.~191.D-0255 indicates that the line formation in the atmosphere
of HD\,54879 can be quite complex.
An anomalous behaviour of weak emission lines in the spectra of the Of?p star NGC\,1624-2 with a 
very strong mean longitudinal magnetic field of the order of 5.35\,kG was
mentioned in the work of \citet{Wade2012}. Opposite to the measurements using absorption
lines, measurements using weak \ion{O}{iii} emission lines yielded
a much lower magnetic field strength of about 2.58\,kG.
The inspection of the high-resolution HARPS\-pol spectra of HD\,54879 reveals the presence of 
weak emission lines belonging 
to \ion{C}{ii}, \ion{O}{ii}, and \ion{Fe}{iii}. Several weak emission lines could, however, not be 
identified using the Vienna Atomic Line Database. A few examples of identified emission lines are 
presented in Fig.~\ref{fig:emis}.
Due to the relatively low $S/N$, of the order of 200,
in the available Stokes~$I$ spectra and the weakness of the corresponding Zeeman features, it is currently not 
possible to conclude on the strength and polarity of the corresponding longitudinal magnetic field.
In Fig.~\ref{fig:weak}, we display the strongest Zeeman feature identified in our spectra, which is 
related to an unidentified emission line with an unknown Land\'e factor at a wavelength close to 
5220.45\,\AA{}.
If this feature is real, then the inferred longitudinal magnetic field would have positive polarity,
which would be inconsistent with the negative polarity measured from the absorption lines 
($\langle B_z \rangle=-578\pm21$\,G; \citealt{Jarvinen2017}).
Obviously, high-resolution spectra obtained at higher S/N are necessary to investigate  in detail 
the emitting regions in HD\,54879.

\cite{Castro2015} infer a dipole magnetic field of 2\,kG from an observed field strength of $-600$\,G.  
It is, however, not certain that the surface magnetic field of HD\,54879 is dipolar. If we assume 
large magnetic spots, then the dipole field should be substantially smaller and a large fraction of 
the surface flux will not couple to the stellar wind and form closed magnetic loops instead.  
The magnetic field strength in sunspots is of the order of kG, but the large-scale surface 
magnetic field is only of the order of G.
The large-scale magnetic field inside the solar convection zone is believed 
to be much stronger, though, comparable with the magnetic field strength in sunspots. Its configuration is 
definitely more complicated than a magnetic dipole (e.g.\ \citealt{Charbonneau2014}). It seems 
plausible that any possible dynamo action is located somewhere near the stellar surface and the 
dynamo-generated magnetic field should be more similar to the magnetic configuration in the solar interior 
than the configuration at the solar surface.
It is therefore possible that the impact of the magnetic field on the stellar wind is 
weaker than one would expect from a dipole field of 2\,kG.

In the following, we will discuss a more sophisticated scenario to explain the HD\,54879 observations.
First of all, 
it is possible that current dynamo models slightly underestimate the intensity of dynamo drivers and thus 
their ability to produce a solar-type dynamo in the convective envelope of HD\,54879 (see the discussion of the 
problem of estimating the intensity of stellar dynamo drivers in \citealt{Shulyak2015}).
Dynamo models usually deal with a more or less 
typical star of a given spectral type while the physical properties of a particular star can deviate from 
standard stellar models, e.g.\ because the star belongs to a binary system. This may explain why the observed 
phenomenon is not common for all O-type stars. \citet{Moss2002} studied an $\alpha^2$-dynamo with a 
non-axisymmetric distribution of the $\alpha$-effect caused by tidal forces in a close binary system. 
This type of dynamo produces large surface spots through the large-scale field, i.e.\ the radial component 
of the mean field is strongly concentrated at certain longitudes on the stellar surface. Adding 
differential rotation then causes the magnetic field to oscillate, but without the migration 
pattern shown by $\alpha^2$-dynamos with axisymmetric $\alpha$-effect. While this scenario creates 
surface spots without requiring flux tubes being brought to the surface by buyoancy, the assumption 
of a strong deviation from axisymmetry in the $\alpha$-effect seems far-fetched in the context of HD\,54879.

Even if the solar-type dynamo in HD\,54879 remains subcritical and dynamo drivers such as differential rotation 
and mirror asymmetric convection produce a more complicated magnetic field than in the solar case, the 
dynamo-generated field configuration can be rather simple, for example of quadrupolar symmetry (this option is 
discussed for stellar dynamos by e.g.\ \citealt{Moss2008}). Stellar winds 
associated with a quadrupole magnetic field have not attracted much attention so far and at least some 
research is required here to rule them out.

In principle, dynamo drivers in HD\,54879 may be insufficiently intensive to produce an oscillating magnetic 
field from a weak seed field, yet sufficient to maintain an already existing magnetic field. Such 
hysteresis dynamos are discussed by \citet{Karak2015}. Fossil magnetic fields as presumed in O-type stars by 
\citet{Moss2003} may provide the initial magnetic field required for hysteresis effects. Such a scenario has 
been discussed by \cite{Singh2017} in a different context. Small-scale dynamo action which 
provides substantial magnetic flux and is modulated by a subcritical oscillating solar-type magnetic 
configuration also seems to be an option that deserves investigation (cf.\ \citealt{Yushkov2018}).       

In the following subsections we consider the impact of a dipole field on the stellar wind, the available X-ray
observations,
and discuss other probable scenarios in an attempt to interpret our observations.

\subsection{Magnetic field impact on the stellar wind}

The impact of a dipole magnetic field rooted in the star on the stellar wind can be 
characterised by the confinement parameter

\begin{equation}
  \eta_*=\frac{B^2_{\rm eq}R^2_*}{\dot{M}v_\infty},
\end{equation}

\noindent
\citep{ud-Doula2002} where $B_{\rm eq}$ is the magnetic field strength at the equator, 
$R_*$ the stellar radius, $\dot{M}$ the mass loss rate, and $v_\infty$ the terminal wind 
speed.  \cite{ud-Doula2008} found a strong reduction of the mass loss rate for the case of 
strong confinement, when magnetic field lines originating close to the magnetic equator remain closed 
and the outflowing gas will thus be trapped at these latitudes. In the equatorial plane magnetic 
field lines will remain closed up to the confinement radius

\begin{equation}
  R_c \approx R_*+0.7(R_A-R_*),
\end{equation}

\noindent
where $R_A$ is the Alfv\'en radius

\begin{equation}
  R_A \approx \left(0.3+0.7 \eta^{1/4}_*\right)R_*.
\end{equation}

\noindent
Magnetic field lines that originate at  angles larger than $\theta_c$ given by

\begin{equation}
   \sin \theta_c = \sqrt{R_*/R_c}
\end{equation}   

\noindent
from the magnetic poles remain closed. Gas can only escape through caps around the magnetic poles 
where open field lines originate. The mass loss rate is reduced by the factor

\begin{equation}
   \frac{\dot{M_B}}{\dot{M_{B=0}}}  \approx 1- \sqrt{1-R_*/R_c}
\end{equation}

\noindent
relative to the case where the magnetic field is absent. 

\begin{table}
\caption{Values of the confinement parameter and the reduction of the mass loss rate }
\begin{center}
\begin{tabular}{cccc}
\hline
$B_* $ & $\eta_*$  & $R_c/R_\odot$ & $\dot{M}/\dot{M_0}$ \\
(G) &  &  &  \\
\hline
100  & 42 & 1.757 &0.3436 \\
 300 & 380 & 2.674&0.2088 \\
 800 & $2.7\times10^3$&4.042 & 0.1325 \\
 2000 & $1.7 \times 10^4$ & 6.105&0.08556 \\
\hline
 \end{tabular}
\end{center}
\label{tab:mdot}
\end{table}

\begin{table}
\caption{Stellar parameters adopted for the numerical simulations.}
\begin{center}
\begin{tabular}{ccc}
$T_*$ 		& (${\rm kK}$) 			& 30.5 \\ 
$\log L$	& ($L_\odot$) 			& 4.42 \\
$R_*$ 		& ($R_\odot$) 			& 6.1 \\
$M_*$ 		& ($M_\odot$) 			& 14 \\
$\log \dot{M_0}$ & ($M_\odot\,{\rm yr}^-1)$ 	& $-9$ \\
$\log \dot{M_B}$ & ($M_\odot\,{\rm yr}^{-1}$) 	& $-10.2$ \\
$v_\infty$ 	& (km\,s$^{-1}$) 		& 1700 \\
\end{tabular}
\label{tab:tab1}
\end{center}
\end{table}

Table~\ref{tab:mdot} lists the confinement parameter, confinement radius, and mass loss rate 
for various values of the magnetic field strength. The effect of stronger fields is a larger extent 
of the magnetically-dominated region, as indicated by the confinement radius and a further reduction of the mass 
loss rate. The stellar parameters have been adopted from \cite{Shenar2017} and are listed in Table \ref{tab:tab1}. 

To illustrate this reduction in the mass loss rate and the distribution of the trapped gas, we have 
carried out a series of numerical simulations using the Nirvana MHD code \citep{Ziegler2011}.  The setup is similar to that 
described in \citet{Kueker2017}. The simulation box is a spherical shell with the stellar
surface as the inner boundary. The outer boundary is typically at a radius of 10 stellar radii. 
The magnetic field is kept fixed on the inner boundary and can evolve elsewhere. We set a fixed mass 
density on the lower boundary and let the radial velocity float. At the outer boundary, the Nirvana 
code's built-in outflow conditions are used. Likewise built-in boundary conditions are used on the symmetry axis.
 
\begin{figure}
\includegraphics[width=0.235\textwidth]{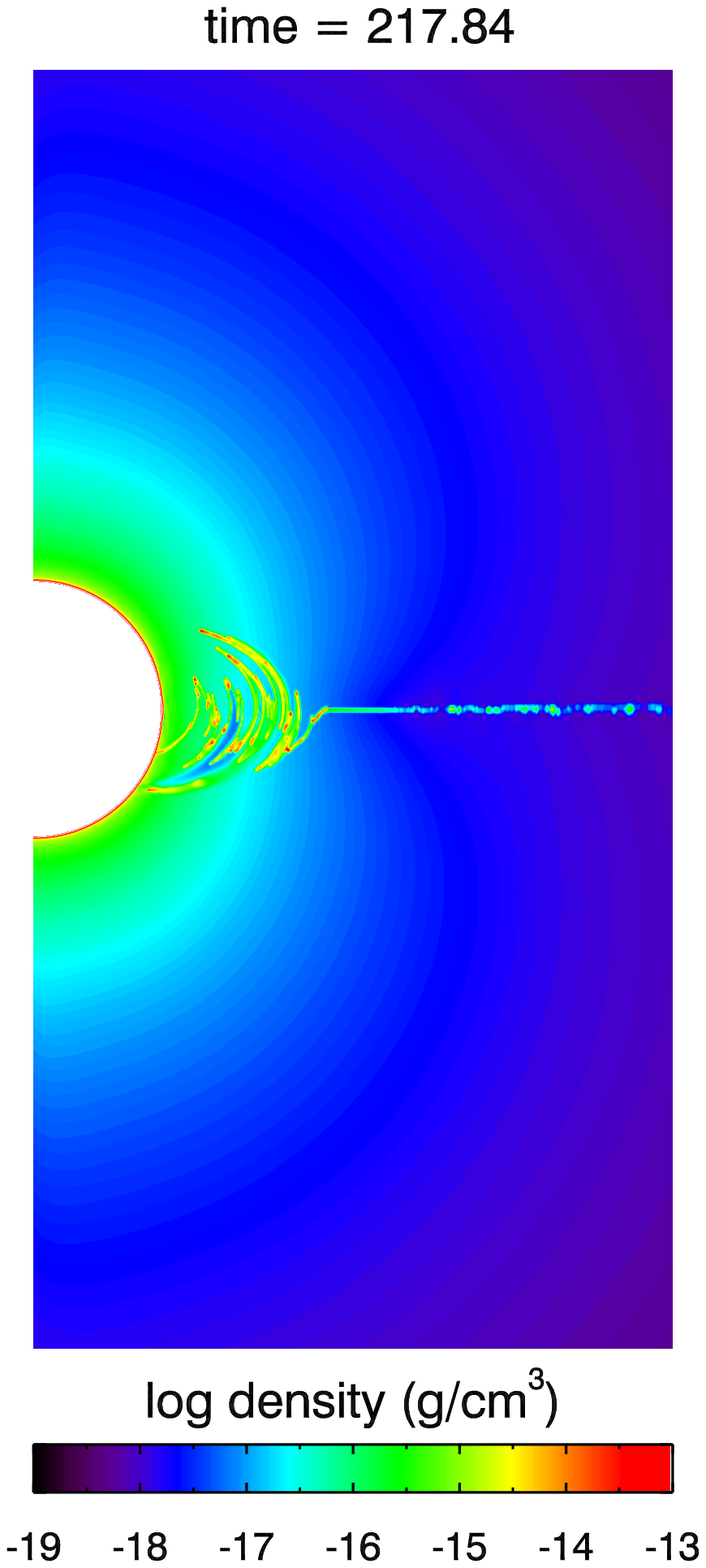}
\includegraphics[width=0.235\textwidth]{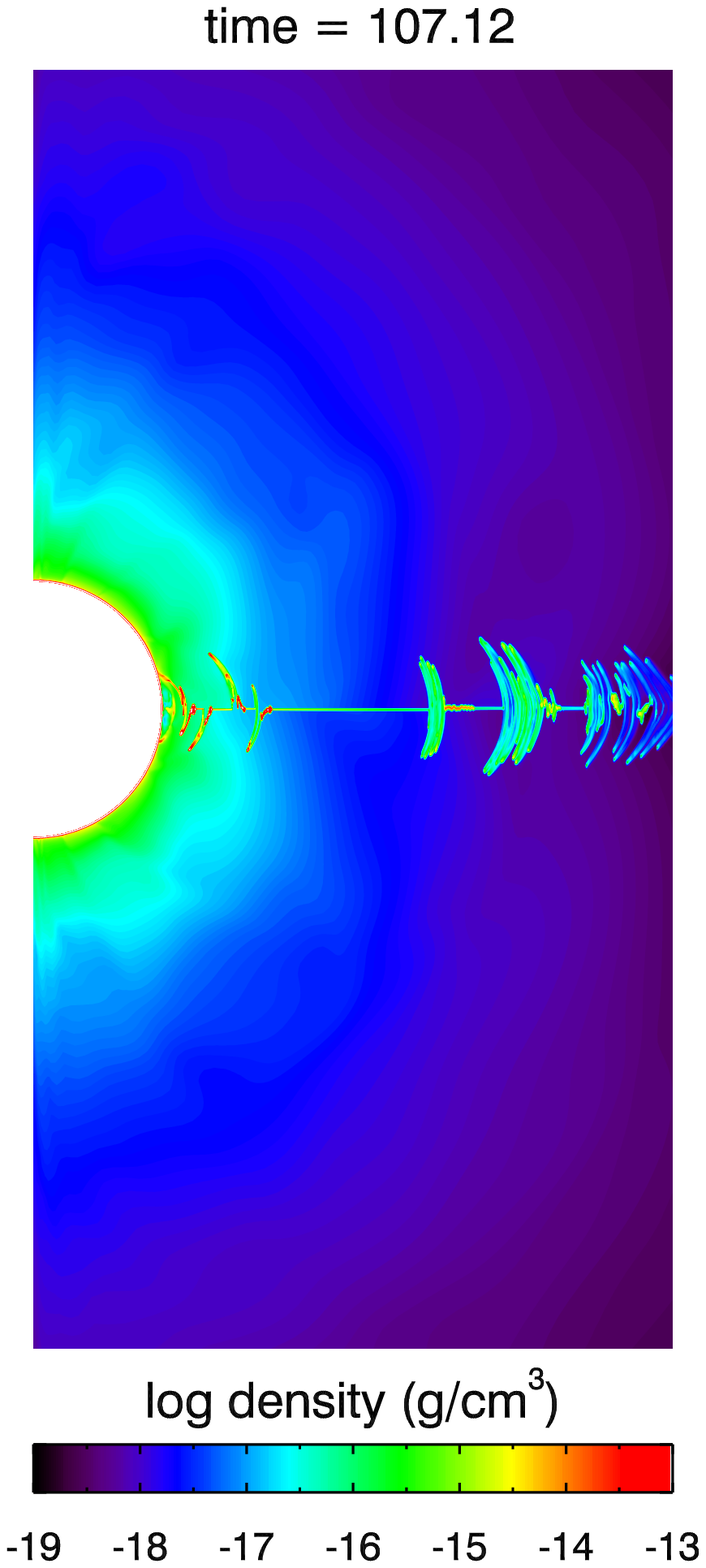}
\caption{
Snapshots from simulations of a line-driven wind and a dipole magnetic field rooted in the star.
The polar field strength is 100\,G (left) and 300\,G (right).
Colours indicate mass density.
}
\label{fig:density}
\end{figure}

 As simulations for polar field strengths of 800\,G and 2\,kG proved to be very time consuming, we studied the first 
two cases listed in Table~\ref{tab:mdot}, namely 100\,G and 300\,G.  Fig.~\ref{fig:density} shows  snapshots of the 
mass density. As indicated by the values of 42 and 380 for the confinement parameter, the magnetic field has a 
profound impact on the gas flow. The 100\,G case shows a fully-developed magnetosphere while in the 300\,G case the 
magnetosphere is still forming. The circular cutout at the left margins of the plots marks the star. 
The left margin is the dipole axis of the magnetic field. Stellar rotation was not included, hence the dipole axis 
of the magnetic field defines the symmetry. Outside the confinement radius $R_c$, gas 
still concentrates in elongated structures along the magnetic field lines, but is ejected away from the 
star in episodic outbursts.

While for very weak magnetic fields -- a small value of the confinement parameter -- the mass loss is steady, 
our simulations show episodic outbreaks of gas previously trapped near the magnetic equator. 
These mass ejections are typical for the case of strong confinement, which is fulfilled even for the weakest field of 100\,G
considered here. During the outbursts, which last about a day, clumps of gas move away from the star in the equatorial 
plane of the magnetic field at high speed. The contribution of these outbursts to the total mass loss rate reaches values 
of the same order as that of the steady wind component, i.e. the total mass loss rate during the outburst can be twice as 
high as during the quiet phase in between. This is in contrast to cases of weak confinement, $\eta_*<1$, where a steady 
disk forms and the mass loss rate  shows little variation with time. 
Obviously, the interaction of the stellar magnetic field with the wind does not explain the drop in
effective temperature and in radial velocity observed during the episode of a strong surface magnetic field.  

\subsection{X-ray spectral analysis}

As our MHD model is isothermal, we cannot make direct predictions regarding the X-ray emission that is 
expected from the material trapped inside the confinement radius. 
It is obvious, however, that X-ray radiation originates from areas near the equatorial plane of the magnetic field.
\citet{Gagne2005} carried out numerical simulations for the star $\theta^1$\,Ori\,C in which the 
energy equation was included. With a cooling function adopted from \citet{MacDonald1981}, they 
found that shock heating would bring the compressed material to temperatures of up to 30\,MK. 
The emission measure calculated from the distributions of mass density and temperature matched 
those observed by the
High Energy Transmission Grating spectrometer (HEGT) onboard the {\em Chandra} space observatory.

A single observation in the X-ray region of HD\,54879 was acquired by {\em XMM-Newton} with an exposure time of about 40\,ks 
(ObsId = ''0780180101'') on 2016 May~1. A previous analysis of these observations was published by \citet{Shenar2017}
who concluded that the X-ray spectrum can be fitted with a thermal model accounting for either two or three components, 
       and implies a higher-than-average X-ray luminosity ($\log L_\text{X} / L_\text{bol} = -6.0$). 
Assuming the three-component model, their result indicates an X-ray temperature reaching values of up 
to $T_\text{X}\approx 20\,$MK. 

\begin{table*}
\begin{center}
 \caption{
X-ray spectral parameters derived from the {\em XMM-Newton} observations of HD\,54879 
assuming different plasma models. The first column lists the applied models, followed by 
the hydrogen column density, the plasma temperatures, the photon index $\Gamma$, and the element abundances.
The indicator of the approximation accuracy $\chi^{2}$ and the degrees of freedom 
are presented in the last column.
The three models with missing entries did not produce any reasonable fitting parameters.
} 
 \label{tab:fitting}
  \begin{tabular}{lccccccc} 
  \hline 
   Model & $N_{H}$ & $kT_{1}$ & $kT_{2}$ & $kT_{3}$ & $\Gamma$ & Abundance & $\chi^{2}$ \\ 
   & ($10^{22} cm^{-2}) $ & (keV) & (keV) & (keV) & (rel.\ un.) & (rel.\ un.) & (d.o.f) \\
  \hline
  \multicolumn{8}{c}{\emph{PN}} \\
  \hline
  WABS$\cdot$APEC & $0.13^{+0.02}_{-0.02}$ & $0.76^{+0.03}_{-0.03}$ & - & - & - & $0.20^{+0.05}_{-0.04}$ & 1.61 (107) \\[5pt]
  WABS$\cdot$(APEC+APEC) & $0.34^{+0.12}_{-0.09}$ & $0.74^{+0.04}_{-0.24}$ & $0.18^{-0.03}_{+0.03}$ & - & - & $0.58^{+1.72}_{-0.24}$ & 1.23 (105) \\[5pt]
  WABS$\cdot$(APEC+APEC+APEC) & $0.50^{+0.10}_{-0.19}$ & $0.11^{+0.07}_{-0.05}$ & $0.28^{+0.80}_{-0.06}$ & $0.77^{+0.09}_{-0.07}$ & - & $\leq$0.63 & 1.17 (103) \\[5pt]
  WABS$\cdot$(APEC+PL) & - & - & - & - & - & - & - \\[5pt]
  WABS$\cdot$(APEC+APEC+PL) & $0.30^{+0.03}_{-0.17}$ & $0.35^{+0.05}_{-0.04}$ & $0.86^{+0.29}_{-0.05}$ & - & $3.98^{+0.37}_{-0.37}$ & $1.90^{+3.10}_{-0.50}$ & 1.47 (103) \\[5pt]
  WABS$\cdot$MEKAL & $0.17^{+0.03}_{-0.02}$ & $0.61^{+0.03}_{-0.03}$ & - & - & - & $0.18^{+0.05}_{-0.04}$ & 1.46 (107) \\ [5pt]
  WABS$\cdot$(MEKAL+MEKAL) & $0.33^{+0.09}_{-0.08}$ & $0.18^{+0.03}_{-0.03}$ & $0.60^{+0.04}_{-0.04}$ & - & - & $0.43^{+0.57}_{-0.16}$ & 1.23 (105) \\
  \hline
  \multicolumn{8}{c}{\emph{MOS1+MOS2}} \\
  \hline
  WABS$\cdot$APEC & $0.14^{+0.03}_{-0.03}$ & $0.74^{+0.04}_{-0.04}$ & - & - & - & $0.14^{+0.04}_{-0.03}$ & 1.07 (118) \\[5pt]
  WABS$\cdot$(APEC+APEC) & $0.26^{+0.11}_{-0.09}$ & $0.75^{+0.08}_{-0.05}$ & $0.21^{+0.11}_{0.05}$ & - & - & $0.18^{+0.12}_{-0.06}$ & 1.02 (116) \\[5pt]
  WABS$\cdot$(APEC+APEC+APEC) & - & - & - & - & - & - & - \\[5pt]
  WABS$\cdot$(APEC+PL) & $0.64^{+0.06}_{-0.19}$ & $0.24^{+0.11}_{-0.02}$ & - & - & $3.70^{+2.02}_{-0.55}$ & $2.02^{+3.08}_{-1.66}$ & 1.30 (116) \\[5pt]
  WABS$\cdot$(APEC+APEC+PL) & - & - & - & - & - & - & - \\[5pt]
  WABS$\cdot$MEKAL & $0.18^{+0.03}_{-0.03}$ & $0.60^{+0.03}_{-0.03}$ & - & - & - & $0.12^{+0.03}_{-0.03}$ & 1.05 (118) \\ [5pt]
  WABS$\cdot$(MEKAL+MEKAL) & $0.27^{+0.11}_{-0.08}$ & $0.20^{+0.10}_{-0.05}$ & $0.60^{+0.05}_{-0.04}$ & - & - & $0.17^{+0.10}_{-0.05}$ & 1.02 (116) \\
  \hline
 \end{tabular}
\end{center}
\end{table*}

\begin{figure*}
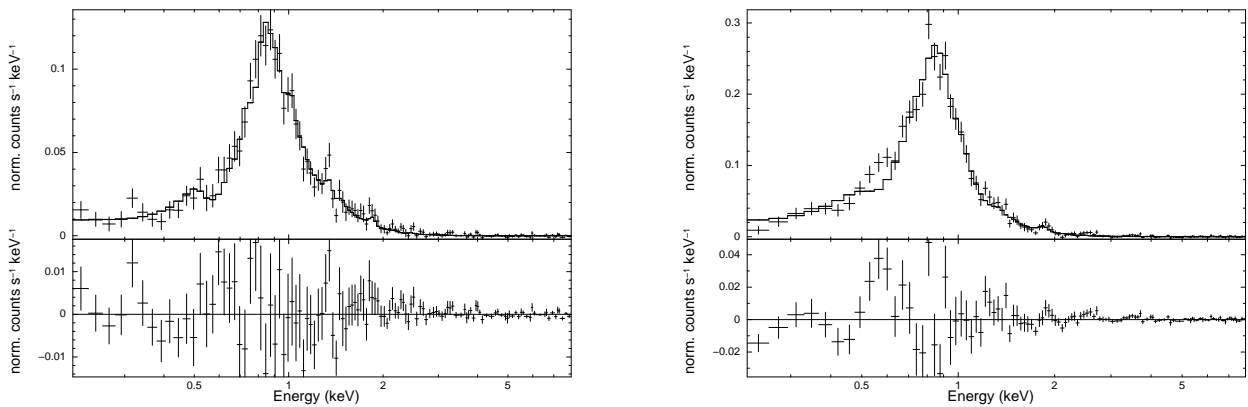

\includegraphics[angle=270,width=0.495\textwidth]{MOS12_mekal.ps}
\includegraphics[angle=270,width=0.495\textwidth]{PN_mekal.ps}
\caption{{\it Left panel:} 
The MOS spectrum (crosses) together with the model WABS$\cdot$MEKAL (solid line).
{\it Right panel:}
The PN spectrum (crosses) together with the model WABS$\cdot$MEKAL (solid line).
The bottom panels show the residuals of the fit.
}
\label{fig:fit}
\end{figure*}

We re-analysed the available {\em XMM-Newton} data using the SAS~17.0 software,
following the recommendations of the SAS team\footnote{\url{https://www.cosmos.esa.int/web/xmm-newton/sas-threads}}.
We modeled the band 0.2-8\,keV of the EPIC-PN spectrum
and EPIC-MOS1+EPIC-MOS2 spectra using the {\sc xspec} 11.0 package. 
The spectra were fitted by a composition of one or two temperature thermal plasma models (APEC or MEKAL) 
and a nonthermal power law (PL) model, which describes synchrotron emission. 
All model compositions were multiplied by the photoelectric 
absorption model WABS for taking into account interstellar absorption.
The best fits are listed in Table~\ref{tab:fitting} and examples 
of the spectral modeling are shown in Fig.~\ref{fig:fit}.

Our results presented in Table~\ref{tab:fitting} indicate that 
the X-ray spectrum of HD\,54879 can be described by thermal plasma models, in agreement with the results 
of \citet{Shenar2017}.
However, our fits are more accurate. 
We note that the spectrum can also be accurately fitted by the sum of APEC and PL models.
A PL model would imply that hard X-rays arise from relativistic electrons
generated through an acceleration mechanism involving strong shocks.
In this model, we would expect a value close to 1.5 for the photon index.
However, since we found much higher values for the photon index, this scenario is less likely.

The RGS1$+$RGS2 spectrum of HD\,54879 was not considered in the study of \citet{Shenar2017}. 
In the spectral band of 6-38\,\AA{} we identified several lines using an approximation of the line profiles
by a sum of Gauss or Gauss-like functions.
This method is described in Section\,3 of \citet{Ryspaeva2018}.
The identified lines were used to calculate the spectral hardness as a ratio of fluxes in lines of adjacent ions.
To calculate the hardness ratios, we used the following lines of ions that are widespread in the spectra of O-stars:
Fe~XXII/Fe~XXI, Fe~XXI/Fe~XX, Fe~XX/Fe~XIX, Fe~XIX/Fe~XVIII, Fe~XVIII/Fe~XVII, Fe~XVII/Fe~XVI,
Ca~XV/Ca~XIV, Ca~XIV/Ca~XIII, Ca~XIII/Ca~XII, Ca~XII/Ca~XI,
Ar~XV/Ar~XIV, and Ar~XIV/Ar~XIII.
The resulting hardness ratios were found in the range $9.8\times10^{-2}$ to $1.46\times10^{1}$ relative units and are 
comparable to those of other O-type  stars presented in Figs.~5 and 6 in the work of \citet{Ryspaeva2018}.  
To conclude, the results of our analysis of the {\em XMM-Newton} observations are not significantly 
different compared to the analysis of other O-type stars.

\subsection{Other scenarios}

For the rapid changes of the magnetic field strength the appearance and decay of magnetic spots seems 
more likely than a variation of a dipole field.
Previous spectroscopic studies of massive stars revealed a variety of spectral variability:
discrete absorption components (DACs) in UV P-Cygni
profiles, optical line profile variability, radial velocity modulations, etc.
Spectroscopic time series observations 
frequently show different absorption features propagating with different accelerations at the same time.
Such features are thought to be signatures of corotating interaction regions (CIRs) in the winds
\citep{Mullan1984}. 
While DACs are ubiquitous among O-type stars, only for a number of massive stars
the current analysis of dynamic UV spectra securely establishes that the variability 
has a photospheric origin and does not originate near the stellar surface
(e.g.\ \citealt{Massa2015}). 
For stars showing CIRs, it has been known for a long time
that the absorption features seen in dynamic spectra must be huge in order to remain in the
line of sight for days.
This means that the linear size of the region responsible
for the excess absorption must be at least 15--20 per cent of the stellar diameter and, in some cases, 
considerably larger.
According to \citet{Cantiello2011a},
convection zones could also be responsible for generating sub-surface
magnetic fields via dynamo action and thereby can account for localized corotating magnetic structures.
They proposed a dynamo in the iron convection 
zones (FeCZ) of OB stars as a mechanism for the generation of surface magnetic fields brought to the 
surface by buoyancy. While the gas density in the  FeCZ is small, the convection velocities are large enough 
to sustain magnetic fields of the observed strength. 
The effect of convection zones in massive stars is expected to increase with increasing luminosity
and lower effective temperature \citep{Cantiello2009}. 
The nature of the dynamo remains unclear, though. 

\citet{Cantiello2011b} carried out 3D MHD simulations that showed the generation of a magnetic 
field, but rotation and shear were required to produce a large-scale magnetic field. 
Bipolar groups of sunspots are believed to be created by the buoyant rise of toroidal flux tubes, which 
are generated by rotational shear.  The slow rotation, however, makes this mechanism unlikely for HD\,54879. 
While a different spot formation mechanism is possible, it is doubtful that a dynamo in the FeCZ of the star 
could generate a large-scale magnetic field. The situation could be different in the convective core, which 
might rotate much faster than the envelope. Rapid rotation would not only favour the generation of a 
large-scale magnetic field in the core, the shear between core and envelope would then generate strong 
toroidal fields, which could rise to the surface and form magnetic spots there.

One can speculate that a short increase of the longitudinal magnetic field
as observed in HD\,54879 may support the scenario of sudden magnetic field ejections into the stellar wind
that is used to explain the bright X-ray flares in supergiant fast X-ray transients.
\citet{Shakura2014} proposed a mechanism to explain in these binaries the observed sporadic
X-ray flares. It is based on an instability of the 
quasi-spherical shell around the magnetosphere of a slowly rotating neutron star (NS).
The authors suggest that the
instability can be triggered by the sudden increase in the mass accretion rate through the magnetosphere due to reconnection of the
large-scale magnetic field sporadically carried by the stellar wind
of the optical OB-companion. The bright
flares due to the proposed mechanism can occur on top of a smooth
variation of mass accretion rate due to, for example, orbital motion
of the NS in a binary system.

\smallskip

\smallskip

To summarize, our results presenting strong magnetic and spectral variability detected during a very short time 
interval clearly show that careful spectropolarimetric monitoring even of stars with very long rotation
periods is necessary to better understand the processes taking place in massive stars. 
Additional X-ray observations are in  particular important to fully characterize the wind 
material confined by the magnetic field of HD\,54879. While the discussed scenarios may have some potential 
in clarifying the sudden increase
of the longitudinal magnetic field, they are currently unable to fully explain the observed phenomena.
The role of an invisible companion is also not clear, as there is no working theoretical 
prediction on how an invisible companion could 
affect the magnetic field strength and the spectral appearance of a massive O-type star just for a very short 
time interval.

%-------------------------------------------------------------------

\section*{Acknowledgments}
We thank the anonymous referee for useful comments.
Further, we would like to thank Drs.~L.~Sidoli and K.~A.~Postnov for fruitful discussions.  
AFK acknowledges support under RSF grant 18-12-00423.
DDS is grateful to RFBR for financial support under grant 18-02-00085. 
Based on observations made with ESO Telescopes at the La Silla Paranal
Observatory under programme IDs~191.D-0255 and 0100.D-0110.
This work has made use of the
VALD database, operated at Uppsala University, the Institute
of Astronomy RAS, Moscow, and the University of Vienna.

%%%%%%%%%%%%%%%%%%%%%%%%%%%%%%%%%%%%%%%%%%%%%%%%%%

%%%%%%%%%%%%%%%%%%%% REFERENCES %%%%%%%%%%%%%%%%%%

\label{lastpage}

\end{document}